# Bits from Photons: Oversampled Image Acquisition Using Binary Poisson Statistics

Feng Yang, *Student Member, IEEE,* Yue M. Lu, *Member, IEEE,*

Luciano Sbaiz, *Member, IEEE,* and Martin Vetterli, *Fellow, IEEE*

## Abstract

We study a new image sensor that is reminiscent of traditional photographic film. Each pixel in the sensor has a binary response, giving only a one-bit quantized measurement of the local light intensity. To analyze its performance, we formulate the oversampled binary sensing scheme as a parameter estimation problem based on quantized Poisson statistics. We show that, with a single-photon quantization threshold and large oversampling factors, the Cramér-Rao lower bound (CRLB) of the estimation variance approaches that of an ideal unquantized sensor, that is, as if there were no quantization in the sensor measurements. Furthermore, the CRLB is shown to be asymptotically achievable by the maximum likelihood estimator (MLE). By showing that the log-likelihood function of our problem is concave, we guarantee the global optimality of iterative algorithms in finding the MLE. Numerical results on both synthetic data and images taken by a prototype sensor verify our theoretical analysis and demonstrate the effectiveness of our image reconstruction algorithm. They also suggest the potential application of the oversampled binary sensing scheme in high dynamic range photography.

F. Yang is with the School of Computer and Communication Sciences, École Polytechnique Fédérale de Lausanne (EPFL), CH-1015 Lausanne, Switzerland (e-mail: feng.yang@epfl.ch; WWW: http://lcav.epfl.ch/people/feng.yang).

Y. M. Lu is with the School of Engineering and Applied Sciences, Harvard University, Cambridge, MA 02138, USA (e-mail: yuelu@seas.harvard.edu; WWW: http://lu.seas.harvard.edu).

L. Sbaiz is with Google Zurich, CH-8002 Zurich, Switzerland (e-mail: luciano.sbaiz@gmail.com).

M. Vetterli is with the School of Computer and Communication Sciences, École Polytechnique Fédérale de Lausanne (EPFL), CH-1015, Lausanne, Switzerland (e-mail: martin.vetterli@epfl.ch; WWW: http://lcav.epfl.ch/people/martin.vetterli).

This work was supported by the Swiss National Science Foundation under grant 200020-132730, and by the European Research Council under grant ERC-247006.

## I. INTRODUCTION

Before the advent of digital image sensors, photography, for the most part of its history, used film to record light information. At the heart of every photographic film are a large number of light-sensitive grains of silver-halide crystals [1]. During exposure, each micron-sized grain has a binary fate: Either it is struck by some incident photons and becomes "exposed", or it is missed by the photon bombardment and remains "unexposed". In the subsequent film development process, exposed grains, due to their altered chemical properties, are converted to silver metal, contributing to opaque spots on the film; unexposed grains are washed away in a chemical bath, leaving behind them transparent regions on the film. Thus, in essence, photographic film is a binary imaging medium, using local densities of opaque silver grains to encode the original light intensity information. Thanks to the small size and large number of these grains, one hardly notices this quantized nature of film when viewing it at a distance, observing only a continuous gray tone.

In this work, we study a new digital image sensor that is reminiscent of photographic film. Each pixel in the sensor has a binary response, giving only a one-bit quantized measurement of the local light intensity. At the start of the exposure period, all pixels are set to $0$. A pixel is then set to $1$ if the number of photons reaching it during the exposure is at least equal to a given threshold $q$. One way to build such binary sensors is to modify standard memory chip technology, where each memory bit cell is designed to be sensitive to visible light [2]. With current CMOS technology, the level of integration of such systems can exceed $10^9 \sim 10^{10}$ (*i.e.*, 1 giga to 10 giga) pixels per chip. In this case, the corresponding pixel sizes (around 50 nm [3]) are far below the diffraction limit of light (see Section II for more details), and thus the image sensor is *oversampling* the optical resolution of the light field. Intuitively, one can exploit this spatial redundancy to compensate for the information loss due to one-bit quantizations, as is classic in oversampled analog-to-digital (A/D) conversions [4]–[7].

Building a binary sensor that emulates the photographic film process was first envisioned by Fossum [8], who coined the name *"digital film sensor"*. The original motivation was mainly out of technical necessity. The miniaturization of camera systems calls for the continuous shrinking of pixel sizes. At a certain point, however, the limited full-well capacity (*i.e.*, the maximum photon-electrons a pixel can hold) of small pixels becomes a bottleneck, yielding very low signal-to-noise ratios (SNRs) and poor dynamic ranges. In contrast, a binary sensor whose pixels only need to detect a few photon-electrons around a small threshold $q$ has much less requirement for full-well capacities, allowing pixel sizes to shrink further.

In this paper, we present a theoretical analysis of the performance of the binary image sensor, and propose an efficient and optimal algorithm to reconstruct images from the binary sensor measurements. Our analysis and numerical simulations demonstrate that the dynamic ranges of the binary sensors can be orders of magnitude higher than those of conventional image sensors, thus providing one more motivation for considering this binary sensing scheme.

Since photon arrivals at each pixel can be well-approximated by a Poisson random process whose rate is determined by the local light intensity, we formulate the binary sensing and subsequent image reconstruction as a parameter estimation problem based on quantized Poisson statistics. Image estimation from Poisson statistics has been extensively studied in the past, with applications in biomedical and astrophysical imaging. Previous work in the literature has used linear models [9], multiscale models [10], [11], and nonlinear piecewise smooth models [12], [13] to describe the underlying images, leading to different (penalized) maximum likelihood and/or Bayesian reconstruction algorithms. The main difference between our work and previous works is that we have only access to one-bit quantized Poisson statistics. The binary quantization and spatial oversampling in the sensing scheme add interesting dimensions to the original problem. As we will show in Section III, the performance of the binary sensor depends on the intricate interplay of three parameters: the average light intensity, the quantization threshold $q$, and the oversampling factor.

The binary sensing scheme studied in this paper also bears resemblance to oversampled A/D conversion schemes with quantizations (see, *e.g.*, [4]–[7]). Previous work on one-bit A/D conversions considers bandlimited signals or, in general, signals living in the range space of some overcomplete representations. The effect of quantization is often approximated by additive noise, which is then mitigated through noise shaping [4], [6], or dithering [7], followed by linear reconstruction. In our work, the binary sensor measurements are modeled as one-bit quantized versions of correlated Poisson random variables (instead of deterministic signals), and we directly solve the statistical inverse problem by using maximum likelihood estimation, without any additive noise approximation.

The rest of the paper is organized as follows. After a precise description of the binary sensing model in Section II, we present three main contributions in this paper:

1. *Estimation performance:* In Section III, we analyze the performance of the proposed binary sensor in estimating a piecewise constant light intensity function. In what might be viewed as a surprising result, we show that, when the quantization threshold $q = 1$ and with large oversampling factors, the Cramér-Rao lower bound (CRLB) [14] of the estimation variance approaches that of unquantized Poisson intensity estimation, that is, as if there were no quantization in the sensor measurements. Furthermore, the CRLB can be asymptotically achieved by a maximum likelihood estimator (MLE), for large oversampling factors. Combined, these two results establish the feasibility of trading spatial resolutions for higher quantization bit depth.

2. *Advantage over traditional sensors:* We compare the oversampled binary sensing scheme with traditional image sensors in Section III-C. Our analysis shows that, with sufficiently large oversampling factors, the new binary sensor can have higher dynamic ranges, making it particularly attractive in acquiring scenes containing both bright and dark regions.

3. *Image reconstruction:* Section IV presents an MLE-based algorithm to reconstruct the light intensity field from the binary sensor measurements. As an important result in this work, we show that the log-likelihood function in our problem is always concave for arbitrary linear field models, thus ensuring the achievement of global optimal solutions by iterative algorithms. For numerically solving the MLE, we present a gradient method, and derive efficient implementations based on fast signal processing algorithms in the polyphase domain [15], [16]. This attention to computational efficiency is important in practice, due to extremely large spatial resolutions of the binary sensors.

Section V presents numerical results on both synthetic data and images taken by a prototype device [17]. These results verify our theoretical analysis on the binary sensing scheme, demonstrate the effectiveness of our image reconstruction algorithm, and showcase the benefit of using the new binary sensor in acquiring scenes with high dynamic ranges.

To simplify the presentation we base our discussions on a one-dimensional (1-D) sensor array, but all the results can be easily extended to the 2-D case.

## II. Imaging by Oversampled Binary Sensors

### A. Diffraction Limit and Linear Light Field Models

In this section, we describe the binary imaging scheme studied in this work. Consider a simplified camera model shown in Fig. 1(a). We denote by $\lambda_0(x)$ the incoming light intensity field (*i.e.*, the radiance map). By assuming that light intensities remain constant within a short exposure period, we model the field as only a function of the spatial variable $x$. Without loss of generality, we assume that the dimension of the sensor array is of one spatial unit, *i.e.*, $0 \leq x \leq 1$.

After passing through the optical system, the original light field $\lambda_0(x)$ gets filtered by the lens, which acts like a linear system with a given impulse response. Due to imperfections (*e.g.*, aberrations) in the lens, the impulse response, a.k.a. the point spread function (PSF) of the optical system, cannot be a Dirac delta, thus, imposing a limit on the resolution of the observable light field. However, a more fundamental physical limit is due to light diffraction [18]. As a result, even if the lens is ideal, the PSF is still unavoidably a small blurry spot [see, for example, Fig. 1(b)]. In optics, such diffraction-limited spot is often called the Airy disk [18], whose radius $R_a$ can be computed as

$$R_a = 1.22 \, wf,$$

where $w$ is the wavelength of the light and $f$ is the F-number of the optical system.

*Example 1:* At wavelength $w = 420$ nm (*i.e.*, for blue visible light) and $f = 2.8$, the radius of the Airy disk is $1.43 \, \mu$m. Two objects with distance smaller than $R_a$ cannot be clearly separated by the imaging system as their Airy disks on the image sensor start blurring together. Current CMOS technology can already make standard pixels smaller than $R_a$, reaching sizes ranging from $0.5$ $\mu$m to $0.7$ $\mu$m [19]. In the case of binary sensors, the simplicity of each pixel allows the feature size to be further reduced. For example, based on standard memory technology, each memory bit-cell (*i.e.*, pixel) can have sizes around $50$ nm [3], making it possible to substantially oversample the light field.

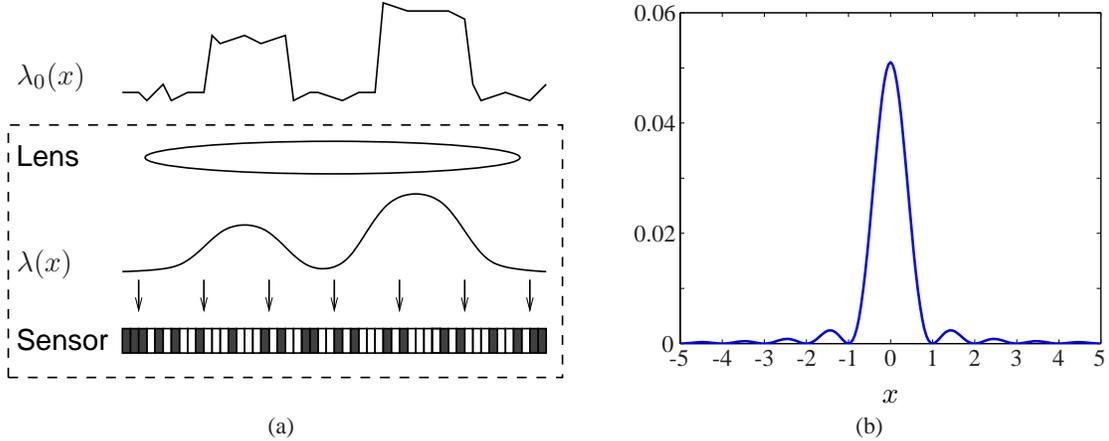

(a)                                                    (b)

Fig. 1.   The imaging model. (a) The simplified architecture of a diffraction-limited imaging system. Incident light field $\lambda_0(x)$ passes through an optical lens, which acts like a linear system with a diffraction-limited point spread function (PSF). The result is a smoothed light field $\lambda(x)$, which is subsequently captured by the image sensor. (b) The PSF (Airy disk) of an ideal lens with a circular aperture.

In what follows, we denote by $\lambda(x)$ the diffraction-limited (*i.e.*, "observable") light intensity field, which is the outcome of passing the original light field $\lambda_0(x)$ through the lens. Due to the lowpass (smoothing) nature of the PSF, the resulting $\lambda(x)$ has a finite spatial-resolution, *i.e.*, it has a finite number of degrees of freedom per unit space.

*Definition 1 (Linear field model):*   In this work, we model the diffraction-limited light intensity field as

$$\lambda(x) = \frac{N}{\tau} \sum_{n=0}^{N-1} c_n \, \varphi(Nx - n),$$

(1)

where $\varphi(x)$ is a nonnegative interpolation kernel, $N$ is a given integer, $\tau$ is the exposure time, and $\{c_n : c_n \geq 0\}$ is a set of free variables.

*Remark 1:*   The constant $N/\tau$ in front of the summation is not essential, but its inclusion here leads to simpler expressions in our later analysis.

The function $\lambda(x)$ as defined in (1) has $N$ degrees of freedom. To guarantee that the resulting light fields are physically meaningful, we require both the interpolation kernel $\varphi(x)$ and the expansion coefficients

$\{c_n\}$ to be nonnegative. Some examples of the interpolation kernels $\varphi(x)$ include the box function,

$$\beta(x) \stackrel{\text{def}}{=} \begin{cases} 1, & \text{if } 0 \leq x \leq 1; \\ 0, & \text{otherwise}, \end{cases} \qquad (2)$$

cardinal B-splines [20],

$$\beta_k(x) = \left(\underbrace{\beta * \ldots * \beta}_{(k+1) \text{ times}}\right)\left(x + \frac{k}{2}\right), \qquad (3)$$

and the squared sinc function, $\sin^2\left(\pi(x - \frac{1}{2})\right) / \left(\pi(x - \frac{1}{2})\right)^2$.

### B. Sampling the Light Intensity Field

The image sensor in Fig. 1(a) works as a sampling device of the light intensity field $\lambda(x)$. Suppose that the sensor consists of $M$ pixels per unit space, and that the $m$th pixel covers the area between $[\frac{m}{M}, \frac{m+1}{M}]$, for $0 \leq m < M$. We denote by $s_m$ the total light exposure accumulated on the surface area of the $m$th pixel within an exposure time period $[0, \tau]$. Then,

$$\begin{aligned} s_m &\stackrel{\text{def}}{=} \int_0^\tau \int_{m/M}^{(m+1)/M} \lambda(x)\, dx\, dt \\ &= \tau \langle \lambda(x), \beta(Mx - m) \rangle, \end{aligned} \qquad (4)$$

where $\beta(x)$ is the box function defined in (2) and $\langle \cdot, \cdot \rangle$ represents the standard $L^2$-inner product. Substitute the light field model (1) into the above equality,

$$\begin{aligned} s_m &= \tau \langle \frac{N}{\tau} \sum_n c_n\, \varphi(Nx - n), \beta(Mx - m) \rangle \\ &= \sum_n c_n \langle N\varphi(Nx - n), \beta(Mx - m) \rangle \\ &= \sum_n c_n \langle \varphi(x), \beta\left(\frac{M(x + n)}{N} - m\right) \rangle, \end{aligned} \qquad (5)$$

where (5) is obtained through a change of variables $(Nx - n) \rightarrow x$.

*Definition 2:* The *spatial oversampling factor*, denoted by $K$, is the ratio between the number of pixels per unit space and the number of degrees of freedom needed to specify the light field $\lambda(x)$ in (1), *i.e.*,

$$K \stackrel{\text{def}}{=} \frac{M}{N}. \qquad (6)$$

In this work, we are interested in the "oversampled" case where $K > 1$. Furthermore, we assume that $K$ is an integer for simplicity of notation. Using (6), and by introducing a discrete filter

$$g_m \stackrel{\text{def}}{=} \langle \varphi(x), \beta(Kx - m) \rangle, \qquad m \in \mathbb{Z}, \qquad (7)$$

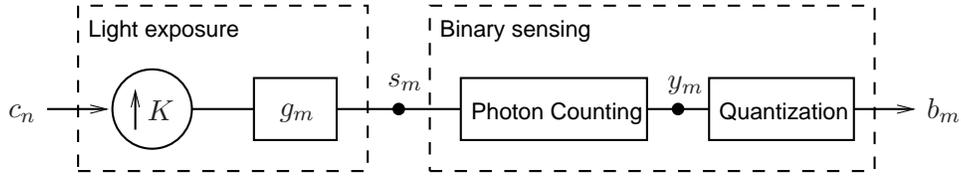

Fig. 2. The signal processing block diagram of the imaging model studied in this paper. In the first step, the light exposure value $s_m$ at the $m$th pixel is related to the expansion coefficients $c_n$ through a concatenation of upsampling and filtering operations. Subsequently, the image sensor converts $\{s_m\}$ into quantized measurements $\{b_m\}$ (see Fig. 3 and the discussions in Section II-C for details of this second step).

we can simplify (5) as

$$s_m = \sum_n c_n\, g_{m-Kn}. \tag{8}$$

The above equality specifies a simple linear mapping from the expansion coefficients $\{c_n\}$ of the light field to the light exposure values $\{s_m\}$ accumulated by the image sensor. Readers familiar with multirate signal processing [15], [16] will immediately recognize that the relation in (8) can be implemented via a concatenation of upsampling and filtering, as shown in the left part of Fig. 2. This observation can also be verified by expressing (8) in the $z$-transform domain

$$S(z) = \sum_n c_n z^{-Kn} G(z) = C(z^K)G(z), \tag{9}$$

and using the fact that $C(z^K)$ is the $z$-transform of the $K$-fold upsampled version of $c_n$. In Section IV, we will further study the signal processing block diagram in Fig. 2 to derive efficient implementations of the proposed image reconstruction algorithm.

*Example 2:* The discrete filter $g_m$ is completely specified by the interpolation kernel $\varphi(x)$ and the oversampling factor $K$. As a simple case, when the kernel $\varphi(x) = \beta(x)$, we can compute from (7) that

$$g_m = \begin{cases} 1/K, & \text{for } 0 \le m < K; \\ 0, & \text{otherwise.} \end{cases} \tag{10}$$

### C. Binary Sensing and One-Bit Poisson Statistics

Fig. 3 illustrates the binary sensor model. Recall from (4) that $\{s_m\}$ denote the exposure values accumulated by the sensor pixels. Depending on the local values of $\{s_m\}$, each pixel (depicted as "buckets" in the figure) collects a different number of photons hitting on its surface. In what follows, we denote by $y_m$ the number of photons impinging on the surface of the $m$th pixel during an exposure period $[0, \tau]$. The relation between $s_m$ and the photon count $y_m$ is stochastic. More specifically, $y_m$ can be modeled as realizations

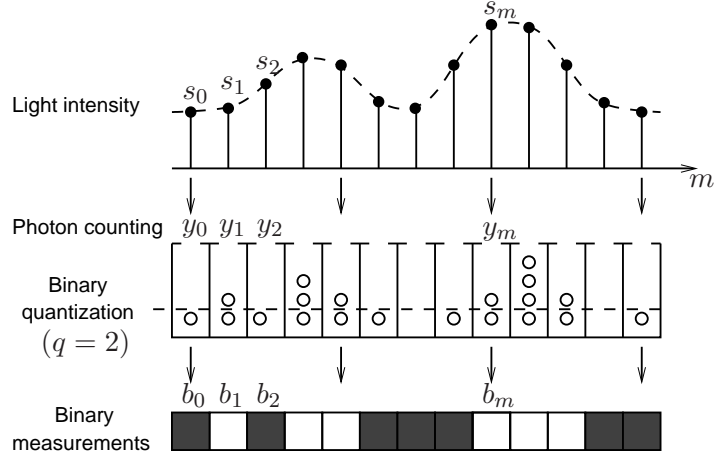

Fig. 3. The model of the binary image sensor. The pixels (shown as "buckets") collect photons, the numbers of which are compared against a quantization threshold $q$. In the figure, we illustrate the case when $q = 2$. The pixel outputs are binary: $b_m = 1$ (*i.e.*, white pixels) if there are at least two photons received by the pixel; otherwise, $b_m = 0$ (*i.e.*, gray pixels).

of a Poisson random variable $Y_m$, whose intensity parameter is equal to $s_m$, *i.e.*,

$$\mathbb{P}(Y_m = y_m; s_m) = \frac{s_m^{y_m} e^{-s_m}}{y_m!}, \qquad \text{for } y_m \in \mathbb{Z}^+ \cup \{0\}. \tag{11}$$

It is a well-known property of the Poisson process that $\mathbb{E}[Y_m] = s_m$. Thus, the average number of photons captured by a given pixel is equal to the local light exposure $s_m$.

As a photosensitive device, each pixel in the image sensor converts photons to electrical signals, whose amplitude is proportional to the number of photons impinging on that pixel.[1] In a conventional sensor design, the analog electrical signals are then quantized by an A/D converter into 8 to 14 bits (usually the more bits the better). In this work, we study a new sensor design using the following binary (*i.e.*, one-bit) quantization scheme.

*Definition 3 (Binary Quantization):* Let $q \geq 1$ be an integer threshold. A binary quantizer is a mapping $Q : \mathbb{Z}^+ \cup \{0\} \longmapsto \{0, 1\}$, such that

$$Q(y) = \begin{cases} 1, & \text{if } y \geq q; \\ 0, & \text{otherwise.} \end{cases}$$

In Fig. 3, we illustrate the binary quantization scheme. White pixels in the figure show $Q(y_m) = 1$ and gray pixels show $Q(y_m) = 0$. We denote by $b_m \stackrel{\text{def}}{=} Q(y_m)$, $b_m \in \{0, 1\}$, the quantized output of the $m$th pixel. Since the photon counts $\{y_m\}$ are drawn from random variables $\{Y_m\}$, so are the binary sensor output $\{b_m\}$, from the random variables $\left\{ B_m \overset{\text{def}}{=} Q(Y_m) \right\}$. Introducing two functions

$$p_0(s) \overset{\text{def}}{=} \sum_{k=0}^{q-1} \frac{s^k}{k!} e^{-s}, \qquad p_1(s) \overset{\text{def}}{=} 1 - \sum_{k=0}^{q-1} \frac{s^k}{k!} e^{-s} \tag{12}$$

we can write

$$\mathbb{P}(B_m = b_m; s_m) = p_{b_m}(s_m), \qquad b_m \in \{0, 1\}. \tag{13}$$

---

[1] The exact ratio between these two quantities is determined by the *quantum efficiency* of the sensor.

*Remark 2:* The noise model considered in this paper is that of Poisson noise. In practice, the performance of image sensors is also influenced by thermal noise, which in our case can be modeled as random bit-flipping in the binary sensor measurements. Due to space constraints, we leave further discussions on this additional noise source and its impact on reconstruction performance to a follow-up work.

### D. Multiple Exposures and Temporal Oversampling

Our previous discussions focus on the case of acquiring a single frame of quantized measurements during the exposure time $[0, \tau]$. As an extension, we can consider multiple exposures and acquire $J$ *consecutive* and *independent* frames. The exposure time for each frame is set to $\tau/J$, so that the total acquisition time remains the same as the single exposure case. In what follows, we call $J$ the *temporal oversampling factor*.

As before, we assume that $\tau \ll 1$ and thus light intensities $\lambda(x)$ stay constant within the entire acquisition time $[0, \tau]$. For the $j$th frame ($0 \le j < J$), we denote by $s_{j,m}$ the light exposure at the $m$th pixel. Following the same derivations as in Section II-B, we can show that

$$s_{j,m} = \frac{1}{J} \sum_n c_n \, g_{m-Kn}, \quad \text{for all } j, \tag{14}$$

where $\{c_n\}$ are the expansion coefficients of the light field $\lambda(x)$, and $g_m$ is the discrete filter defined in (7). The only difference between (14) and (8) is the extra factor of $1/J$, due to the change of exposure time from $\tau$ to $\tau/J$. In the $z$-domain, similar to (9),

$$S_j(z) = \frac{1}{J} C(z^K) G(z). \tag{15}$$

In what follows, we establish an equivalence between temporal oversampling and spatial oversampling. More precisely, we will show that an $M$-pixel sensor taking $J$ independent exposures (*i.e.*, with $J$-times oversampling in time) is mathematically equivalent to a single sensor consisting of $MJ$ pixels.

First, we introduce a new sequence $\widetilde{s}_m, 0 \le m < MJ$, constructed by interlacing the $J$ exposure sequences $\{s_{j,m}\}$. For example, when $J = 2$, the new sequence is

$$\boxed{s_{0,0}}, s_{1,0}, \boxed{s_{0,1}}, s_{1,1}, \dots, \boxed{s_{0,m}}, s_{1,m}, \dots, \boxed{s_{0,M-1}}, s_{1,M-1},$$

where $\{s_{0,m}\}$ and $\{s_{1,m}\}$ alternate. In general, $\widetilde{s}_m$ can be obtained as

$$\widetilde{s}_m \overset{\text{def}}{=} s_{j,n}, \qquad m = Jn + j, \ 0 \le j < J, \ 0 \le n < M. \tag{16}$$

In multirate signal processing, the above construction is called the *polyphase representation* [15], [16], and its alternating subsequences $\{s_{j,m}\}_{j=0}^{J-1}$ the *polyphase components*.

*Proposition 1:* Let $\widetilde{g}_m$ be a filter whose $z$-transform

$$\widetilde{G}(z) \overset{\text{def}}{=} G(z^J)(1 + z^{-1} + \ldots + z^{-(J-1)})/J, \tag{17}$$

where $G(z)$ is the $z$-transform of the filter $g_m$ defined in (7). Then,

$$\widetilde{s}_m = \sum_n c_n \, \widetilde{g}_{m-KJn}. \tag{18}$$

*Proof:* See Appendix A. ∎

*Remark 3:* Proposition 1 formally establishes the equivalence between spatial and temporal oversampling. We note that (18) has exactly the same form as (8), and thus the mapping from $\{c_n\}$ to $\{\widetilde{s}_m\}$ can be implemented by the same signal processing operations shown in Fig. 2—we only need to change the upsampling factor from $K$ to $KJ$ and the filter from $g_m$ to $\widetilde{g}_m$. In essence, by taking $J$ consecutive exposures with an $M$-pixel sensor, we get the same light exposure values $\{\widetilde{s}_m\}$, as if we had used a more densely packed sensor with $MJ$ pixels.

*Remark 4:* Taking multiple exposures is a very effective way to increase the total oversampling factor of the binary sensing scheme. The key assumption in our analysis is that, during the $J$ consecutive exposures, the light field remains constant over time. To make sure this assumption holds for arbitrary values of $J$, we set the exposure time for each frame to $\tau/J$, for a fixed and small $\tau$. Consequently, the maximum temporal oversampling factor we can achieve in practice will be limited by the readout speed of the binary sensor.

Thanks to the equivalence between spatial and temporal oversampling, we only need to focus on the single exposure case in our following discussions on the performance of the binary sensor and image reconstruction algorithms. All the results we obtain extend directly to the multiple exposure case.

## III. PERFORMANCE ANALYSIS

In this section, we study the performance of the binary image sensor in estimating light intensity information, analyze the influence of the quantization threshold and oversampling factors, and demonstrate the new sensor's advantage over traditional sensors in terms of higher dynamic ranges. In our analysis, we assume that the light field is piecewise constant, *i.e.*, the interpolation kernel $\varphi(x)$ in (1) is the box function $\beta(x)$. This simplifying assumption allows us to derive closed-form expressions for several important performance measures of interest. Numerical results in Section V suggest that the results and conclusions we obtain in this section applies to the general linear field model in (1) with different interpolation kernels.

### A. The Cramér-Rao Lower Bound (CRLB) of Estimation Variances

From Definition 1, reconstructing the light intensity field $\lambda(x)$ boils down to estimating the unknown deterministic parameters $\{c_n\}$. Input to our estimation problem is a sequence of binary sensor measurements $\{b_m\}$, which are realizations of Bernoulli random variables $\{B_m\}$. The probability distributions of $\{B_m\}$ depend on the light exposure values $\{s_m\}$, as in (13). Finally, the exposure values $\{s_m\}$ are linked to the light intensity parameters $\{c_n\}$ in the form of (8).

Assume that the light field $\lambda(x)$ is piecewise constant. We have computed in Example 2 that, under this case, the discrete filter $g_m$ used in (8) is a constant supported within $[0, K-1]$, as in (10). The mapping (8) between $\{c_n\}$ and $\{s_m\}$ can now be simplified as

$$s_m = c_n/K, \quad \text{for } nK \le m < (n+1)K. \tag{19}$$

We see that the parameters $\{c_n\}$ have disjoint regions of influence, meaning, $c_0$ can only be sensed by a group of pixels $\{s_0, \ldots, s_{K-1}\}$, $c_1$ by $\{s_K, \ldots, s_{2K-1}\}$, and so on. Consequently, the parameters $\{c_n\}$ can be estimated one-by-one, independently of each other.

In what follows, and without loss of generality, we focus on estimating $c_0$ from the block of binary measurements $\boldsymbol{b} \stackrel{\text{def}}{=} [b_0, \ldots, b_{K-1}]^T$. For notational simplicity, we will drop the subscript in $c_0$ and use $c$ instead. To analyze the performance of the binary sensing scheme, we first compute the CRLB [14], which provides a theoretical lower bound on the variance of any unbiased estimator.

Denote by $\mathcal{L}_{\boldsymbol{b}}(c)$ the likelihood function of observing $K$ binary sensor measurement $\boldsymbol{b}$. Then,

$$\mathcal{L}_{\boldsymbol{b}}(c) \stackrel{\text{def}}{=} \mathbb{P}(B_m = b_m, 0 \le m < K; c),$$

$$= \prod_{m=0}^{K-1} \mathbb{P}(B_m = b_m; c), \tag{20}$$

$$= \prod_{m=0}^{K-1} p_{b_m}(c/K), \tag{21}$$

where (20) is due to the independence of the photon counting processes at different pixel locations, and (21) follows from (13) and (19). Defining $K_1$ $(0 \le K_1 < K)$ to be the number of "1"s in the binary sequence $\boldsymbol{b}$, we can simplify (21) as

$$\mathcal{L}_{\boldsymbol{b}}(c) = \big(p_1(c/K)\big)^{K_1} \big(p_0(c/K)\big)^{K-K_1}. \tag{22}$$

*Proposition 2:* The CRLB of estimating the light intensity $c$ from $K$ binary sensor measurements with threshold $q \ge 1$ is

$$\text{CRLB}_{\text{bin}}(K, q) = c \left( \sum_{j=0}^{q-1} \frac{(q-1)!(c/K)^{-j}}{(q-1-j)!} \right) \left( \sum_{j=0}^{\infty} \frac{(q-1)!(c/K)^j}{(q+j)!} \right), \quad \text{for } c > 0. \tag{23}$$

*Proof:* See Appendix B. ∎

It will be interesting to compare the performance of our binary image sensor with that of an ideal sensor which does not use quantization at all. To that end, consider the same situation as before, where we use $K$ pixels to observe a constant light intensity value $c$. The light exposure $s_m$ at each pixel is equal to $c/K$, as in (19). Now, unlike the binary sensor which only takes one-bit measurements, consider an ideal sensor that can perfectly record the number of photon arrivals at each pixel. By referring to Fig. 3, the sensor measurements in this case will be $\{y_m\}$, whose probability distributions are given in (11).

In Appendix C, we compute the CRLB of this unquantized sensing scheme as

$$\text{CRLB}_{\text{ideal}}(K) = c, \tag{24}$$

which is natural and reflects the fact that the variance of a Poisson random variable is equal to its mean (*i.e.*, $c$, in our case).

To be sure, we always have $\text{CRLB}_{\text{bin}}(K, q) > \text{CRLB}_{\text{ideal}}(K)$, for arbitrary oversampling factor $K$ and quantization threshold $q$. This is not surprising, as we lose information by one-bit quantizations. In practice, the ratio between the two CRLBs provides a measure of performance degradations incurred by the binary sensors. What is surprising is that the two quantities can be made arbitrarily close, when $q = 1$ and $K$ is large, as shown by the following proposition.

*Proposition 3:* For $q = 1$,

$$\text{CRLB}_{\text{bin}}(K, q) = c + \frac{c^2}{2K} + \mathcal{O}\left(\frac{1}{K^2}\right), \tag{25}$$

which converges to $\text{CRLB}_{\text{ideal}}(K)$ as the oversampling factor $K$ goes to infinity. For $q \geq 2$,

$$\text{CRLB}_{\text{bin}}(K, q) / \text{CRLB}_{\text{ideal}}(K) > 1.31, \tag{26}$$

and $\lim_{K \to \infty} \text{CRLB}_{\text{bin}}(K, q) / \text{CRLB}_{\text{ideal}}(K) = \infty$.

*Proof:* Specializing the expression (23) for $q = 1$, we get

$$\text{CRLB}_{\text{bin}}(K, 1) = c \left(1 + \frac{c}{2K} + \frac{c^2}{3!K^2} + \frac{c^3}{4!K^3} + \dots\right),$$

and thus (25). The statements for cases when $q \geq 2$ are shown in Appendix D. ∎

Proposition 3 indicates that it is feasible to use oversampling to compensate for information loss due to binary quantizations. It follows from (25) that, with large oversampling factors, the binary sensor operates as if there were no quantization in its measurements. It is also important to note that this desirable tradeoff between spatial resolution and estimation variance only works for a single-photon threshold (*i.e.*, $q = 1$). For other choices of the quantization threshold, the "gap" between $\text{CRLB}_{\text{bin}}(K, q)$ and $\text{CRLB}_{\text{ideal}}(K)$, measured in terms of their ratio, cannot be made arbitrarily small, as shown in (26). In fact, it quickly tends to infinity as the oversampling factor $K$ increases.

The results of Proposition 3 can be intuitively understood as follows: The expected number of photons collected by each pixel during light exposure is equal to $s_m = c/K$. As the oversampling factor $K$ goes to infinity, the mean value of the Poisson distribution tends to zero. Consequently, most pixels on the sensor will only get zero or one photon, with the probability of receiving two or more photons at a pixel close to zero. In this case, with high probability, a binary quantization scheme with threshold $q = 1$ does not lose information. In contrast, if $q \geq 2$, the binary sensor measurements will be almost uniformly zero, making it nearly impossible to differentiate between different light intensities.

### B. Asymptotic Achievability of the CRLB

In what follows, we show that, when $q = 1$, the CRLB derived in (23) can be asymptotically achieved by a simple maximum likelihood estimator (MLE). Given a sequence of $K$ binary measurements $\boldsymbol{b}$, the MLE we seek is the parameter that maximizes the likelihood function $\mathcal{L}_{\boldsymbol{b}}(c)$ in (22). More specifically,

$$\widehat{c}_{\mathrm{ML}}(\boldsymbol{b}) \stackrel{\text{def}}{=} \underset{0 \leq c \leq S}{\arg \max} \ \mathcal{L}_{\boldsymbol{b}}(c)$$

$$= \underset{0 \leq c \leq S}{\arg \max} \ \big(1 - p_0(c/K)\big)^{K_1} \big(p_0(c/K)\big)^{K - K_1}, \tag{27}$$

where we substitute $p_1(c/K)$ in (22) by its equivalent form $1 - p_0(c/K)$. The lower bound of the search domain $c \geq 0$ is chosen according to physical constraints, *i.e.*, the light field can not take negative values. The upper bound $c \leq S$ becomes necessary when $K_1 = K$, in which case the likelihood function $\mathcal{L}_{\boldsymbol{b}}(c) = p_1(c/K)^K$ is monotonically increasing with respect to the light intensity level $c$.

*Lemma 1:* The MLE solution to (27) is

$$\widehat{c}_{\mathrm{ML}}(\boldsymbol{b}) = \begin{cases} K \, p_0^{[-1]}(1 - K_1/K), & \text{if } 0 \leq K_1 \leq K(1 - p_0(S/K)), \\ S, & \text{otherwise,} \end{cases} \tag{28}$$

where $p_0^{[-1]}(x)$ is the inverse function of $p_0(x)$.

*Remark 5:* From the definition in (12), we can easily verify that $\frac{d}{dx} p_0(x) < 0$ for all $x > 0$. It follows that the function $p_0(x)$ is strictly decreasing for $x > 0$ and that the inverse $p_0^{[-1]}(x)$ is well-defined. For example, when $q = 1$, we have $p_0(x) = e^{-x}$ and thus $p_0^{[-1]}(x) = -\log(x)$. In this particular case, and for $K_1 \ll K$, we have $\widehat{c}_{\mathrm{ML}}(\boldsymbol{b}) = -K \log(1 - K_1/K) \approx K_1$. It follows that we can use the sum of the $K$ binary measurements as a first-order approximation of the light intensity estimation.

*Proof:* At the two extreme cases, when $K_1 = 0$ or $K_1 = K$, it is easy to see that (28) is indeed the solution to (27). Next, we assume that $0 < K_1 < K$.

Computing the derivative of $\mathcal{L}_{\boldsymbol{b}}(c)$ and setting it to zero, we can verify that the equation $\frac{d}{dc} \mathcal{L}_{\boldsymbol{b}}(c) = 0$ has a single solution at

$$\widehat{c}_{\max} = K \, p_0^{[-1]}(1 - K_1/K).$$

Since $\mathcal{L}_{\boldsymbol{b}}(c) \geq 0$ and $\mathcal{L}_{\boldsymbol{b}}(0) = \lim_{c \to \infty} \mathcal{L}_{\boldsymbol{b}}(c) = 0$, we conclude that the likelihood function $\mathcal{L}_{\boldsymbol{b}}(c)$ achieves its maximum value at $\widehat{c}_{\max}$. Finally, the MLE solution $\widehat{c}_{\mathrm{ML}} = \min\{\widehat{c}_{\max}, S\}$, and thus, we have (28). ∎

*Theorem 1:* When $q = 1$, we have

$$\mathbb{E}[\widehat{c}_{\mathrm{ML}}(\boldsymbol{b})] = c + \varepsilon_1 + \mathcal{O}(1/K), \quad \text{for } c < S - 2, \tag{29}$$

where $|\varepsilon_1| \leq 2c\,e^{1-c}\left(\frac{ec}{S-1}\right)^{S-1}$. Meanwhile, the mean squared error (MSE) of the estimator approaches $\mathrm{CRLB}_{\mathrm{ideal}}$, *i.e.*,

$$\mathbb{E}\left[(\widehat{c}_{\mathrm{ML}}(\boldsymbol{b}) - c)^2\right] = c + \varepsilon_2 + \mathcal{O}(1/K), \quad \text{for } c < S - 2, \tag{30}$$

where $|\varepsilon_2| \leq 2c(c+1)e^{-c}\left(\frac{ec}{S-2}\right)^{S-2}$.

*Remark 6:* It is easy to verify that, for fixed $c$, the two terms $\varepsilon_1$ and $\varepsilon_2$ converge (very quickly) to $0$ as $S$ tends to infinity. It then follows from (29) and (30) that the MLE is asymptotically unbiased and efficient, in the sense that

$$\lim_{S \to \infty} \lim_{K \to \infty} \mathbb{E}[\widehat{c}_{\mathrm{ML}}(\boldsymbol{b})] = c \quad \text{and} \quad \lim_{S \to \infty} \lim_{K \to \infty} \mathbb{E}\left[(\widehat{c}_{\mathrm{ML}}(\boldsymbol{b}) - c)^2\right] = c.$$

We leave the formal proof of this theorem to Appendix E. Its main idea can be summarized as follows. As $K$ goes to infinity, the area of each pixel tends to zero, so does the average number of photons arriving at that pixel. As a result, most pixels on the sensor will get only zero or one photon during exposure. A single-photon binary quantization scheme can record perfectly the pattens of "0"s and "1"s on the sensor. It loses information only when a pixel receives two or more photons, but the probability of such events tends to zero as $K$ increases.

Suppose, now, that we use a quantization threshold $q \geq 2$. In this case, as $K$ tends to infinity, the binary responses of different pixels will almost always be "0", essentially obfuscating the actual light intensity values. This problem leads to poor performance in the MLE. As stated in the following proposition, the asymptotic MSE for $q \geq 2$ becomes $c^2$ instead of $c$.

*Proposition 4:* When $q \geq 2$, the MLE is asymptotically *biased*, that is, for any fixed $c$ and $S$,

$$\lim_{K \to \infty} \mathbb{E}[\widehat{c}_{\mathrm{ML}}(\boldsymbol{b})] = 0. \tag{31}$$

Meanwhile, the MSE becomes

$$\lim_{K \to \infty} \mathbb{E}\left[(\widehat{c}_{\mathrm{ML}}(\boldsymbol{b}) - c)^2\right] = c^2. \tag{32}$$

*Proof:* See Appendix F. ∎

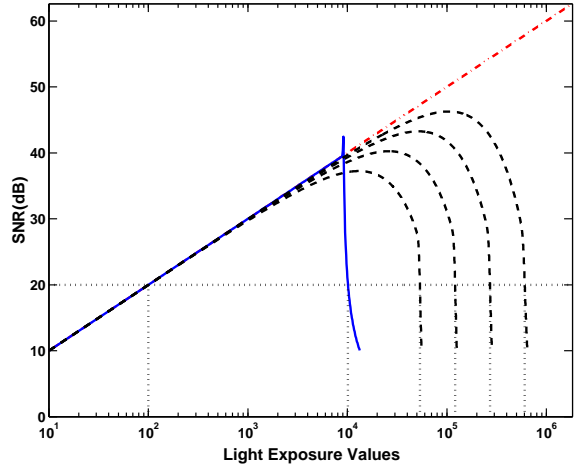

Fig. 4.    Performance comparisons of three different sensing schemes ("BIN", "IDEAL", and "SAT") over a wide range of light exposure values $c$ (shown in logarithmic scale). The dash-dot line (in red) represents the "IDEAL" scheme with no quantization; The solid line (in blue) corresponds to the "SAT" scheme with a saturation point set at $C_{\max} = 9130$ [21]; The four dashed lines (in black) correspond to the "BIN" scheme with $q = 1$ and different oversampling factors (from left to right, $K = 2^{13}$, $2^{14}$, $2^{15}$ and $2^{16}$, respectively).

### C. Advantages over Traditional Sensors

In what follows, we demonstrate the advantage of the oversampled binary sensing scheme, denoted by "BIN", in achieving higher dynamic ranges. We focus on the case where the quantization threshold is set to $q = 1$. For comparisons, we also consider the following two alternative sensing schemes: The first, denoted by "IDEAL", uses a single pixel to estimate the light exposure parameter (*i.e.*, nonoversampled), but that pixel can record perfectly the number of photon arrivals during exposure; The second scheme, denoted by "SAT", is very similar to the first, with the addition of a saturation point $C_{\max}$, beyond which the pixel can hold no more photons. Note that in our discussions, the "SAT" scheme serves as an idealized model of conventional image sensors, for which the saturation is caused by the limited full-well capacity of the semiconductor device. The general trend of conventional image sensor design has been to pack more pixels per chip by reducing pixel sizes, leading to lower full-well capacities and thus lower saturation values.

Fig. 4 compares performances of the three different sensing schemes (*i.e.*, "BIN", "IDEAL", and "SAT") over a wide range of light exposure values. We measure the performances in terms of signal-to-noise ratios (SNRs), defined as

$$\mathrm{SNR} = 10 \log_{10} \frac{c^2}{\mathbb{E}[(\widehat{c} - c)^2]},$$

where $\widehat{c}$ is the estimation of the light exposure value we obtain from each of the sensing schemes.

We observe that the "IDEAL" scheme (the red dash-dot line in the figure) represents an upper-bound of the estimation performance. To see this, denote by $y$ the number of photons that arrive at the pixel during exposure. Then $y$ is a realization of a Poisson random variable $Y$ with intensity equal to the light exposure value $c$, *i.e.*,

$$\mathbb{P}(Y = y; c) = \frac{c^y e^{-c}}{y!}.$$

Maximizing this function over $c$, we can compute the MLE for the "IDEAL" scheme as $\widehat{c}_{\text{IDEAL}}(y) = y$. It is easy to verify that this estimator is unbiased, *i.e.*, $\mathbb{E}[\widehat{c}_{\text{IDEAL}}(Y)] = \mathbb{E}[Y] = c$, and that it achieves the ideal CRLB in (24), *i.e.*, $\text{var}(\widehat{c}_{\text{IDEAL}}(Y)) = \text{var}(Y) = c$. Accordingly, we can compute the SNR as

$$\text{SNR}_{\text{IDEAL}} = 10 \log_{10}(c^2/c) = 10 \log_{10}(c),$$

which appears as a straight line in our figure with the light exposure values $c$ shown in logarithmic scale.

The solid line in the figure corresponds to the "SAT" scheme, with a saturation point set at $C_{\max} = 9130$, which is the full well capacity of the image sensor reported in [21]. The sensor measurement in this case is $y_{\text{SAT}} \stackrel{\text{def}}{=} \min\{y, C_{\max}\}$, and the estimator we use is

$$\widehat{c}_{\text{SAT}}(y_{\text{SAT}}) = y_{\text{SAT}}. \tag{33}$$

We can see that the "SAT" scheme initially has the same performance as "IDEAL". It remains this way until the light exposure value $c$ approaches the saturation point $C_{\max}$, after which there is a drastic drop[2] in SNR. Denoting by $\text{SNR}_{\min}$ the minimum acceptable SNR in a given application, we can then define the dynamic range of a sensor as the range of $c$ for which the sensor achieves at least $\text{SNR}_{\min}$. For example, if we choose $\text{SNR}_{\min} = 20$ dB, then, as shown in the figure, the SAT scheme has a dynamic range from $c = 10^2$ to $c \approx 10^4$, or, if measured in terms of ratios, $100 : 1$.

Finally, the three dashed lines represent the "BIN" scheme with $q = 1$ and increasing oversampling factors (from left to right: $K = 2^{13}$, $2^{14}$, $2^{15}$ and $2^{16}$, respectively). We use the MLE given in (28) and plot the corresponding estimation SNRs. We see that, within a large range of $c$, the performance of the "BIN" scheme is very close to that of the "IDEAL" scheme that does not use quantization. This verifies our analysis in Theorem 1, which states that the "BIN" scheme with a single-photon threshold can approach the ideal unquantized CRLB when the oversampling factor is large enough. Furthermore, when compared with the "SAT" scheme, the "BIN" scheme has a more gradual decrease in SNR when the light exposure values increase, and has a higher dynamic range. For example, when $K = 2^{16}$, the dynamic range of the "BIN" scheme spans from $c = 10^2$ to $c = 10^{5.8}$, about two orders of magnitude higher than that of "SAT". In Section V, we will present a numerical experiment that points to a potential application of the binary sensor in high dynamic range photography.

---

[2]The estimator in (33) is biased around $c = C_{\max}$. For a very narrow range of light intensity values centered around $C_{\max}$, the MSE of this biased estimator is lower than the ideal CRLB. Thus, there is actually a short "spike" in SNR right before the drop.

*Remark 7:* Note that $K$ is the product of the spatial oversampling factor and the temporal oversampling factor. For example, the pixel pitch of the image sensor reported in [21] is $1.65 \mu m$. If the binary sensor is built on memory chip technology, with a pitch size of 50 nm [3], then the maximum spatial oversampling factor is about 1089. To achieve $K = 2^{13}$, $2^{14}$, $2^{15}$ and $2^{16}$, respectively, as required in Fig. 4, we then need to have temporal oversampling factors ranging from 8 to 60. Unlike traditional sensors which require multi-bit quantizers, the binary sensors only need one-bit comparators. This simplicity in hardware can potentially lead to faster readout speeds, making it practical to apply temporal oversampling.

## IV. Optimal Image Reconstruction and Efficient Implementations

In the previous section, we studied the performance of the binary image sensor, and derived the MLE for a piecewise-constant light field model. Our analysis establishes the optimality of the MLE, showing that, with single-photon thresholding and large oversampling factors, the MLE approaches the performance of an ideal sensing scheme without quantization. In this section, we extends the MLE to the general linear field model in (1), with arbitrary interpolation kernels. As a main result of this work, we show that the log-likelihood function is always concave. This desirable property guarantees the global convergence of iterative numerical algorithms in solving the MLE.

### A. Image Reconstruction by MLE

Under the linear field model introduced in Definition 1, reconstructing an image [*i.e.*, the light field $\lambda(x)$] is equivalent to estimating the parameters $\{c_n\}$ in (1). As shown in (8), the light exposure values $\{s_m\}$ at different sensors are related to $\{c_n\}$ through a linear mapping, implemented as upsampling followed by filtering as in Fig. 2. Since it is linear, the mapping (8) can be written as a matrix-vector multiplication

$$\boldsymbol{s} = \boldsymbol{G}\,\boldsymbol{c}, \tag{34}$$

where $\boldsymbol{s} \stackrel{\text{def}}{=} [s_0, s_1, \ldots, s_{M-1}]^T$, $\boldsymbol{c} \stackrel{\text{def}}{=} [c_0, c_1, \ldots, c_{N-1}]^T$, and $\boldsymbol{G}$ is an $M \times N$ matrix representing the combination of upsampling (by $K$) and filtering (by $g_m$). Each element of $\boldsymbol{s}$ can then be written as

$$s_m = \boldsymbol{e}_m^T \boldsymbol{G} \boldsymbol{c}, \tag{35}$$

where $\boldsymbol{e}_m$ is the $m$th standard Euclidean basis vector.[3]

*Remark 8:* In using the above notations, we do not distinguish between single exposure and multiple exposures, whose equivalence has been established by Proposition 1 in Section II-D. In the case of multiple exposures, the essential structure of $\boldsymbol{G}$—upsampling followed by filtering—remains the same. All we need to do is to replace $\boldsymbol{s}$ by the interlaced sequence $\{\widetilde{s}_m\}$ constructed in (16), the oversampling factor $K$ by $KJ$, and the filter $g_m$ by $\widetilde{g}_m$ in (17).

---

[3]Here we use zero-based indexing. Thus, $\boldsymbol{e}_0 \stackrel{\text{def}}{=} [1, 0, \ldots, 0]^T$, $\boldsymbol{e}_1 \stackrel{\text{def}}{=} [0, 1, \ldots, 0]^T$, and so on.

Similar to our derivations in (20) and (21), the likelihood function given $M$ binary measurements $\boldsymbol{b} \overset{\text{def}}{=} [b_0, b_1, \ldots, b_{M-1}]^T$ can be computed as

$$
\begin{aligned}
\mathcal{L}_{\boldsymbol{b}}(\boldsymbol{c}) &= \prod_{m=0}^{M-1} \mathbb{P}(B_m = b_m; s_m) \\
&= \prod_{m=0}^{M-1} p_{b_m}(\boldsymbol{e}_m^T \boldsymbol{G} \boldsymbol{c}),
\end{aligned}
\tag{36}
$$

where (36) follows from (12) and (35). In our subsequent discussions, it is more convenient to work with the log-likelihood function, defined as

$$
l_{\boldsymbol{b}}(\boldsymbol{c}) \overset{\text{def}}{=} \log \mathcal{L}_{\boldsymbol{b}}(\boldsymbol{c}) = \sum_{m=0}^{M-1} \log p_{b_m}(\boldsymbol{e}_m^T \boldsymbol{G} \boldsymbol{c}).
\tag{37}
$$

For any given observation $\boldsymbol{b}$, the MLE we seek is the parameter that maximizes $\mathcal{L}_{\boldsymbol{b}}(\boldsymbol{c})$, or equivalently, $l_{\boldsymbol{b}}(\boldsymbol{c})$. Specifically,

$$
\begin{aligned}
\widehat{\boldsymbol{c}}_{\text{ML}}(\boldsymbol{b}) &\overset{\text{def}}{=} \underset{\boldsymbol{c} \in [0, S]^N}{\arg\max} \; l_{\boldsymbol{b}}(\boldsymbol{c}) \\
&= \underset{\boldsymbol{c} \in [0, S]^N}{\arg\max} \sum_{m=0}^{M-1} \log p_{b_m}(\boldsymbol{e}_m^T \boldsymbol{G} \boldsymbol{c}).
\end{aligned}
\tag{38}
$$

The constraint $\boldsymbol{c} \in [0, S]^N$ means that every parameter $c_n$ should satisfy $0 \le c_n \le S$, for some preset maximum value $S$.

*Example 3:* As discussed in Section III, when the light field is piecewise-constant, different light field parameters $\{c_n\}$ can be estimated independently. In that case, the likelihood function has only one variable [see (22)] and can be easily visualized. In Fig. 5, we plot $\mathcal{L}_{\boldsymbol{b}}(c)$ in (22) and the corresponding log-likelihood function $l_{\boldsymbol{b}}(c)$, under different choices of the quantization thresholds. We observe from the figures that the likelihood functions are not concave, but the log-likelihood functions indeed are. In what follows, we will show that this result is general, namely, the log-likelihood functions in the form of (37) are always concave.

*Lemma 2:* For any two integers $i, j$ such that $0 \le i \le j < \infty$ or $0 \le i < j = \infty$, the function

$$
\log \sum_{k=i}^{j} \frac{x^k e^{-x}}{k!}
$$

is concave on the interval $x \in [0, \infty)$.

*Proof:* See Appendix G. ∎

*Theorem 2:* For arbitrary binary sensor measurements $\boldsymbol{b}$, the log-likelihood function $l_{\boldsymbol{b}}(\boldsymbol{c})$ defined in (37) is concave on the domain $\boldsymbol{c} \in [0, S]^N$.

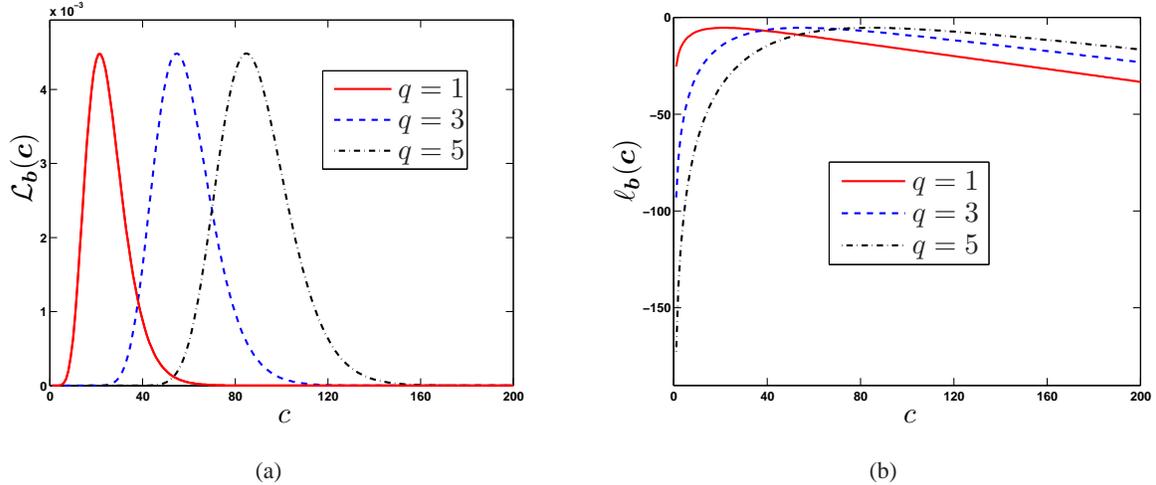

(a)  (b)

Fig. 5. The likelihood and log-likelihood functions for piecewise-constant light fields. (a) The likelihood functions $\mathcal{L}_{\boldsymbol{b}}(c)$, defined in (22), under different choices of the quantization thresholds $q = 1, 3, 5$, respectively. (b) The corresponding log-likelihood functions. In computing these functions, we set the parameters in (22) as follows: $K = 12$, *i.e.*, the sensor is 12-times oversampled. The binary sensor measurements contain 10 "1"s, *i.e.*, $K_1 = 10$.

*Proof:* It follows from the definition in (12) that, for any $b_m \in \{0, 1\}$, the function $\log p_{b_m}(s)$ is either

$$\log \sum_{k=0}^{q-1} \frac{s^k e^{-s}}{k!} \quad \text{or} \quad \log \sum_{k=q}^{\infty} \frac{s^k e^{-s}}{k!}. \tag{39}$$

We can apply Lemma 2 in both cases, and show that $\{\log p_{b_m}(s)\}$ are concave functions for $s \geq 0$. Since the sum of concave functions is still concave and the composition of a concave function with a linear mapping ($s_m = \boldsymbol{e}_m^T \boldsymbol{G} \boldsymbol{c}$) is still concave, we conclude that the log-likelihood function defined in (37) is concave. ∎

In general, there is no closed-form solution to the maximization problem in (38). An MLE solution has to be found through numerical algorithms. Theorem 2 guarantees the global convergence of these iterative numerical methods.

## B. Iterative Algorithm and Efficient Implementations

We compute the numerical solution of the MLE by using a standard gradient ascent method. Denote by $\boldsymbol{c}^{(k)}$ the estimation of the unknown parameter $\boldsymbol{c}$ at the $k$th step. The estimation $\boldsymbol{c}^{(k+1)}$ at the next step is obtained by

$$\boldsymbol{c}^{(k+1)} = P_{\mathcal{D}} \left( \boldsymbol{c}^{(k)} + \gamma_k \nabla l_{\boldsymbol{b}}(\boldsymbol{c}^{(k)}) \right), \tag{40}$$

where $\nabla l_{\boldsymbol{b}}(\boldsymbol{c}^{(k)})$ is the gradient of the log-likelihood function evaluated at $\boldsymbol{c}^{(k)}$, $\gamma_k$ is the step-size at the current iteration, and $P_{\mathcal{D}}$ is the projection onto the search domain $\mathcal{D} \stackrel{\text{def}}{=} [0, S]^N$. We apply $P_{\mathcal{D}}$ to ensure that all estimations of $\boldsymbol{c}$ lie in the search domain.

Taking the derivative of the log-likelihood function $l_{\boldsymbol{b}}(\boldsymbol{c})$ in (37), we can compute the gradient as

$$\nabla l_{\boldsymbol{b}}(\boldsymbol{c}^{(k)}) = \boldsymbol{G}^T \left[ D_{b_0}(s_0^{(k)}), D_{b_1}(s_1^{(k)}), \ldots, D_{b_{M-1}}(s_{M-1}^{(k)}) \right]^T, \tag{41}$$

where $\boldsymbol{s}^{(k)} \stackrel{\text{def}}{=} [s_0^{(k)}, \ldots, s_{M-1}^{(k)}]^T = \boldsymbol{G}\boldsymbol{c}^{(k)}$ is the current estimation of the light exposure values, and

$$D_b(s) \stackrel{\text{def}}{=} \frac{d}{ds} \log p_b(s) \quad \text{for } b = 0, 1.$$

For example, when $q = 1$, we have $p_0(s) = e^{-s}$ and $p_1(s) = 1 - e^{-s}$. In this case, $D_0(s) = -1$ and $D_1(s) = 1/(1 - e^{-s})$, respectively.

The choice of the step size $\gamma_k$ has significant influence over the speed of convergence of the above iterative algorithm. We follow [9] by choosing, at each step, a $\gamma_k$ so that the gradient vectors at the current and the next iterations are approximately orthogonal to each other. By assuming that the estimates $\boldsymbol{s}^{(k+1)}$ and $\boldsymbol{s}^{(k)}$ at consecutive iterations are close to each other, we can use the following first-order approximation

$$D_b(s_m^{(k+1)}) \approx D_b(s_m^{(k)}) + H_b(s_m^{(k)})(s_m^{(k+1)} - s_m^{(k)}),$$

where

$$H_b(s) \stackrel{\text{def}}{=} \frac{d}{ds} D_b(s) = \frac{d^2}{ds^2} \log p_b(s), \quad \text{for } b = 0, 1.$$

It follows that

$$\nabla l_{\boldsymbol{b}}(\boldsymbol{c}^{(k+1)}) = \boldsymbol{G}^T \left[ D_{b_0}(s_0^{(k+1)}), D_{b_1}(s_1^{(k+1)}), \ldots, D_{b_{M-1}}(s_{M-1}^{(k+1)}) \right]$$
$$\approx \nabla l_{\boldsymbol{b}}(\boldsymbol{c}^{(k)}) + \boldsymbol{G}^T \text{diag} \left\{ H_{b_0}(s_0^{(k)}), H_{b_1}(s_1^{(k)}), \ldots, H_{b_{M-1}}(s_{M-1}^{(k)}) \right\} (\boldsymbol{s}^{(k+1)} - \boldsymbol{s}^{(k)}). \tag{42}$$

Assuming that the gradient update $\boldsymbol{c}^{(k)} + \gamma_k \nabla l_{\boldsymbol{b}}(\boldsymbol{c}^{(k)})$ is inside of the constraint set $\mathcal{D}$, we can neglect the projection operator $P_{\mathcal{D}}$ in (40), and write

$$\boldsymbol{s}^{(k+1)} - \boldsymbol{s}^{(k)} = \boldsymbol{G}(\boldsymbol{c}^{(k+1)} - \boldsymbol{c}^{(k)}) = \gamma_k \boldsymbol{G} \nabla l_{\boldsymbol{b}}(\boldsymbol{c}^{(k)}).$$

Substituting the above equality into (42), we get

$$\nabla l_{\boldsymbol{b}}(\boldsymbol{c}^{(k+1)}) \approx \nabla l_{\boldsymbol{b}}(\boldsymbol{c}^{(k)}) + \gamma_k \, \boldsymbol{G}^T \text{diag} \left\{ H_{b_0}(s_0^{(k)}), H_{b_1}(s_1^{(k)}), \ldots, H_{b_{M-1}}(s_{M-1}^{(k)}) \right\} \boldsymbol{G} \nabla l_{\boldsymbol{b}}(\boldsymbol{c}^{(k)}).$$

Finally, by requiring that $\nabla l_{\boldsymbol{b}}(\boldsymbol{c}^{(k+1)})$ be orthogonal to $\nabla l_{\boldsymbol{b}}(\boldsymbol{c}^{(k)})$, we compute the optimal step size as

$$\gamma_k = \frac{\|\nabla l_{\boldsymbol{b}}(\boldsymbol{c}^{(k)})\|^2}{\left\| \text{diag} \left\{ \sqrt{-H_{b_0}(s_0^{(k)})}, \ldots, \sqrt{-H_{b_{M-1}}(s_{M-1}^{(k)})} \right\} \boldsymbol{G} \nabla l_{\boldsymbol{b}}(\boldsymbol{c}^{(k)}) \right\|^2}. \tag{43}$$

*Remark 9:* By definition, $H_b(s)$ (for $b = 0, 1$) are the second-order derivatives of concave functions (see Lemma 2), and are thus nonpositive. Consequently, the terms $\sqrt{-H_b(s)}$ in the denominator of (43) are well-defined.

PSfrag replacements PSfrag replacements PSfrag replacements PSfrag replacements

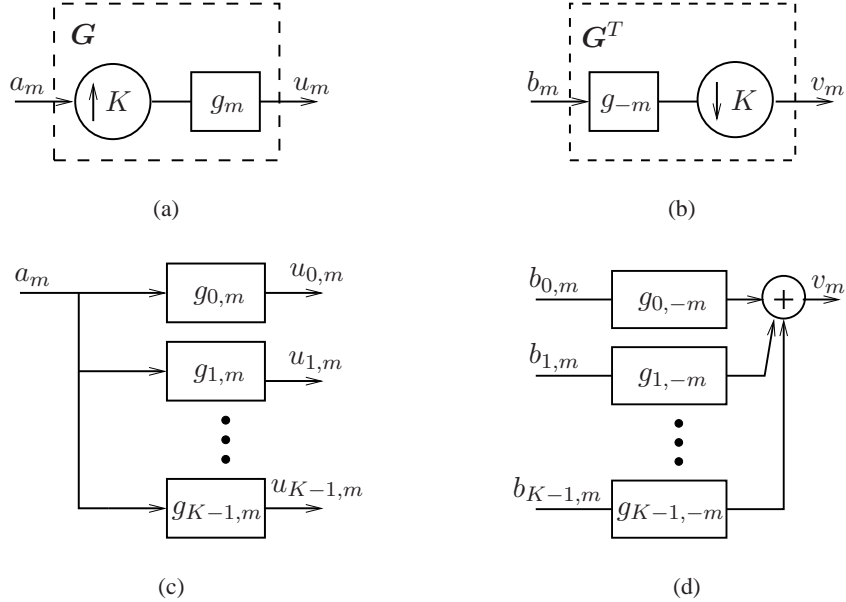

Fig. 6. Signal processing implementations of $\boldsymbol{G}\boldsymbol{a}$ and $\boldsymbol{G}^T\boldsymbol{b}$. (a) The product $\boldsymbol{G}\boldsymbol{a}$ can be obtained by upsampling followed by filtering. (b) The product $\boldsymbol{G}^T\boldsymbol{b}$ can be obtained by filtering followed by downsampling. Note that the filter used in (b) is $g_{-m}$, *i.e.*, the "flipped" version of $g_m$. (c) The polyphase domain implementation of (a). (d) The polyphase domain implementation of (b).

At every iteration of the gradient algorithm, we need to update the gradient and the step size $\gamma_k$. We see from (41) and (43) that the computations always involve matrix-vector products in the form of $\boldsymbol{G}\boldsymbol{a}$ and $\boldsymbol{G}^T\boldsymbol{b}$, for some vectors $\boldsymbol{a}, \boldsymbol{b}$. The matrix $\boldsymbol{G}$ is of size $M \times N$, where $M$ is the total number of pixels. In practice, $M$ will be in the range of $10^9 \sim 10^{10}$ (*i.e.*, gigapixel per chip), making it impossible to directly implement the matrix operations. Fortunately, the matrix $\boldsymbol{G}$ used in both formulae is highly structured, and can be implemented as upsampling followed by filtering (see our discussions in Section II-B and the expression (8) for details). Similarly, the transpose $\boldsymbol{G}^T$ can be implemented by filtering (by $g_{-m}$) followed by downsampling, essentially "flipping" all the operations in $\boldsymbol{G}$. Fig. 6(a) and Fig. 6(b) summarizes these operations.

We note that the implementations illustrated in Fig. 6(a) and Fig. 6(b) are not yet optimized: For example, the input to the filter $g_m$ in Fig. 6(a) is an upsampled sequence, containing mostly zero elements; In Fig. 6(b), we compute a full filtering operation (by $g_{-m}$), only to discard most of the filtering results in the subsequent downsampling step. All these deficiencies can be eliminated by using the tool of polyphase representations from multirate signal processing [15], [16], as follows.

First, we split the filter $g_m$ into $K$ non-overlapping *polyphase components* $g_{0,m}, g_{1,m}, \ldots, g_{K-1,m}$, defined as

$$g_{k,m} = g_{Km+k}, \quad \text{for } 0 \le k < K. \tag{44}$$

Intuitively, the polyphase components specified in (44) are simply downsampled versions of the original filter $g_m$, with the sampling locations of all these polyphase components forming a complete partition. The mapping between the filter $g_m$ and its polyphase components is one-to-one. To reconstruct $g_m$, we can easily verify that, in the $z$-domain,

$$G(z) = G_0(z^K) + z^{-1}G_1(z^K) + \ldots z^{-(K-1)}G_{K-1}(z^K). \tag{45}$$

Following the same steps as above, we can also split the sequences $u_m$ and $b_m$ in Fig. 6 into their respective polyphase components $u_{0,m}, u_{1,m}, \ldots, u_{K-1,m}$ and $b_{0,m}, b_{1,m}, \ldots, b_{K-1,m}$.

*Proposition 5:* Denote by $U_k(z)$ and $B_k(z)$ (for $0 \leq k < K$) the $z$-transforms of $u_{k,m}$ and $b_{k,m}$, respectively. Then,

$$U_k(z) = A(z)G_k(z), \quad \text{for } 0 \leq k < K, \tag{46}$$

and

$$V(z) = \sum_{k=0}^{K-1} B_k(z)G_k(z^{-1}). \tag{47}$$

*Proof:* See Appendix H. ∎

The results of Proposition 5 require some further explanations. What equation (46) suggests is an alternative implementation of $\boldsymbol{Ga}$, as shown in Fig. 6(c). We compute $K$ parallel convolutions between the input $a_m$ and the polyphase filters $\{g_{k,m}\}$. The channel outputs are the polyphase components $\{u_{k,m}\}$, which can be combined to form the desired output $u_m$. Similarly, it follows from (47) that $\boldsymbol{G}^T\boldsymbol{b}$ can be implemented by the parallel filtering scheme in Fig. 6(d).

The new implementations in Fig. 6(c) and Fig. 6(d) are significantly faster than their respective counterparts. To see this, suppose that the filter $g_m$ has $L$ coefficients. It is easy to see that the original implementation in Fig. 6(a) requires $\mathcal{O}(KL)$ arithmetic operations for every pixel in $a_m$. In contrast, each individual channel in Fig. 6(c) requires only $\mathcal{O}(L/K)$ arithmetic operations (due to the shorter supports of the polyphase filters), and thus the total cost of Fig. 6(c) stays at $\mathcal{O}(L)$ operations per pixel. This represents a $K$-fold reduction in computational complexities. A similar analysis also shows that Fig. 6(d) needs $K$-times fewer operations than Fig. 6(b). Recall that $K$ is the oversampling factor of our image sensor. As we operate in highly oversampled regimes (*e.g.*, $K = 1024$) to compensate for information loss due to one-bit quantizations, the above improvements make our algorithms orders of magnitude faster.

## V. NUMERICAL RESULTS

We present several numerical results in this section to verify our theoretical analysis and the effectiveness of the proposed image reconstruction algorithm.

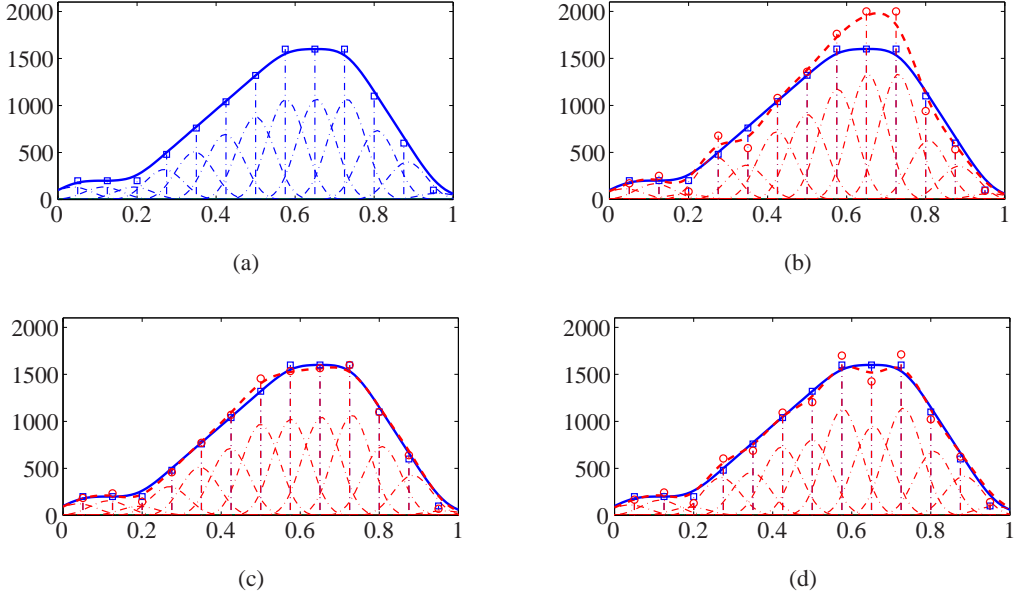

Fig. 7.  Binary sensing and reconstructions of 1-D light fields. (a) The original light field $\lambda(x)$, modeled as a linear combination of shifted spline kernels. (b) The reconstruction result obtained by the proposed MLE-based algorithm, using measurements taken by a sensor with spatial oversampling factor $K = 256$. (c) An improved reconstruction result due to the use of a larger spatial oversampling factor $K = 2048$. (d) An alternative result, obtained by keeping $K = 256$ but taking $J = 8$ consecutive exposures.

### A.  1-D Synthetic Signals

Consider a 1-D light field $\lambda(x)$ shown in Fig. 7(a). The interpolation filter $\varphi(x)$ we use is the cubic B-spline function $\beta_3(x)$ defined in (3). We can see that $\lambda(x)$ is a linear combination of the shifted kernels, with the expansion coefficients $\{c_n\}$ shown as blue dots in the figure.

We simulate a binary sensor with threshold $q = 1$ and oversampling factor $K = 256$. Applying the proposed MLE-based algorithm in Section IV, we obtain a reconstructed light field (the red dashed curve) shown in Fig. 7(b), together with the original "ground truth" (the blue solid curve). We observe that the low-light regions are well-reconstructed but there exist large "overshoots" in the high-light regions.

We can substantially improve the reconstruction quality by increasing the oversampling factor of the sensor. Fig. 7(c) shows the result obtained by increasing the spatial oversampling factor to $K = 2048$. Alternatively, we show in Fig. 7(d) a different reconstruction result obtained by keeping the original spatial oversampling factor at $K = 256$, but taking $J = 8$ consecutive exposures. Visually, the two sensor configurations, *i.e.*, $K = 2048, J = 1$ and $K = 256, J = 8$, lead to very similar reconstruction performances. This observation agrees with our previous theoretical analysis in Section II-D on the equivalence between spatial and temporal oversampling schemes.

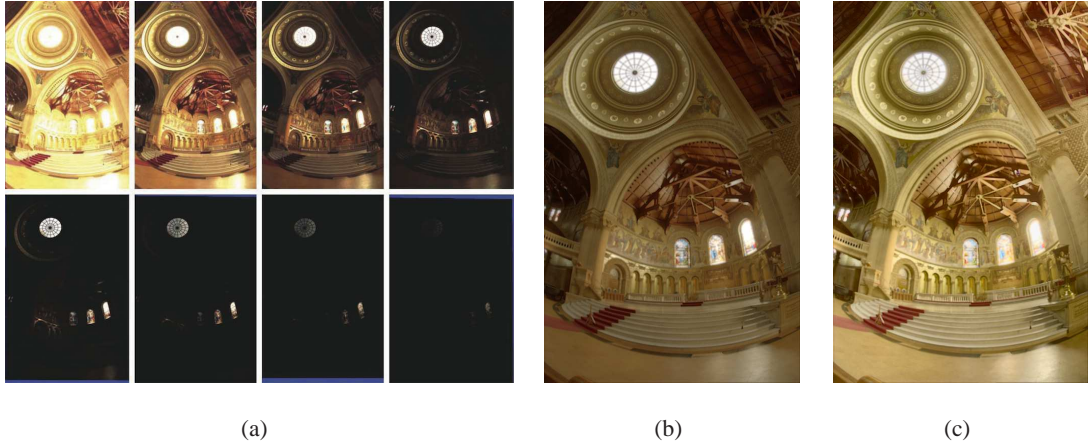

|       |       |       |
|-------|-------|-------|
| (a)   | (b)   | (c)   |

Fig. 8. High dynamic range photography using the binary sensor. (a) A sequence of images taken inside of a church with decreasing exposure times [22]. (b) The reconstructed high dynamic range radiance map (in logarithmic scales) using our MLE reconstruction algorithm. (c) The tone-mapped version of the reconstructed radiance map.

## B. Acquiring Scenes with High Dynamic Ranges

A well-known difficulty in photography is the limited dynamic ranges of the image sensors. Capturing both very bright and very dark regions faithfully in a single image is difficult. For example, Fig. 8(a) shows several images taken inside of a church with different exposure times [22]. The scene contains both sun-lit areas and shadow regions, with the former over a thousand times brighter than the latter. Such high dynamic ranges are well-beyond the capabilities of conventional image sensors. As a result, these images are either overexposed or underexposed, with no single image rendering details in both areas. In light of this problem, an active area of research in computational photography is to reconstruct a high dynamic range radiance map by combining multiple images with different exposure settings (see, *e.g.*, [22], [23]). While producing successful results, such multi-exposure approaches can be time-consuming.

In Section III-C, we have shown that the binary sensor studied in this work can achieve higher dynamic ranges than conventional image sensors. To demonstrate this advantage, we use the high dynamic range radiance map obtained in [22] as the ground truth data [*i.e.*, the light field $\lambda(x)$ as defined in (1)], and simulate the acquisition of this scene by using a binary sensor with a single photon threshold. The spatial oversampling factor of the binary sensor is set to $32 \times 32$, and the temporal oversampling factor is 256 (*i.e.*, 256 independent frames). Similar to our previous experiment on 1-D signals, we use a cubic B-spline kernel [*i.e.*, $\varphi(x) = \beta_3(x)$] along each of the spatial dimensions. Fig. 8(b) shows the reconstructed radiance map using our algorithm described in Section IV. Since the radiance map has a dynamic range of $3.3 \times 10^5 : 1$, the image is shown in logarithmic scale. To have a visually more pleasing result, we also shown in Fig. 8(c) a tone-mapped [23] version of the reconstruction. We can see from Fig. 8(b) and Fig. 8(c) that details in both light and shadow regions have been faithfully preserved in the reconstructed radiance map, suggesting

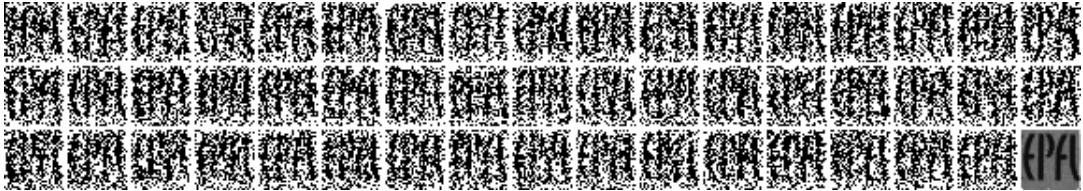

Fig. 9.   Image reconstruction from the binary measurements taken by a SPAD sensor [17], with a spatial resolution of $32 \times 32$ pixels. The final image (lower-right corner) is obtained by incorporating 4096 consecutive frames, 50 of which are shown in the figure.

the potential application of the binary sensor in high dynamic range photography.

### C. Results on Real Sensor Data

We have also applied our reconstruction algorithm to images taken by an experimental sensor based on single photon avalanche diodes (SPADs) [17]. The sensor has binary-valued pixels with single-photon sensitivities, *i.e.*, the quantization threshold is $q = 1$. Due to its experimental nature, the sensor has limited spatial resolution, containing an array of only $32 \times 32$ detectors. To emulate the effect of spatial oversampling, we apply temporal oversampling and acquire 4096 independent binary frames of a static scene. In this case, we can estimate the light intensity at each pixel independently by using the closed-form MLE solution in (28). Fig. 9 shows 50 such binary images, together with the final reconstruction result (at the lower-right corner). The quality of reconstruction verifies our theoretical model and analysis.

## VI. Conclusions

We presented a theoretical study of a new image sensor that acquires light information using one-bit pixels—a scheme reminiscent of traditional photographic film. By formulating the binary sensing scheme as a parameter estimation problem based on quantized Poisson statistics, we analyzed the performance of the binary sensor in acquiring light intensity information. Our analysis shows that, with a single-photon quantization threshold and large oversampling factors, the binary sensor performs much like an ideal sensor, as if there were no quantization. To recover the light field from binary sensor measurements, we proposed an MLE-based image reconstruction algorithm. We showed that the corresponding log-likelihood function is always concave, thus guaranteeing the global convergence of numerical solutions. To solve for the MLE, we adopt a standard gradient method, and derive efficient implementations using fast signal processing algorithms in the polyphase domain. Finally, we presented numerical results on both synthetic data and images taken by a prototype sensor. These results verify our theoretical analysis and demonstrate the effectiveness of our image reconstruction algorithm. They also point to the potential of the new binary sensor in high dynamic range photography applications.

APPENDIX

### A. Proof of Proposition 1

The sequence $\widetilde{s}_m$ in (16) can be written, equivalently, as $\widetilde{s}_m = \sum_{j=0}^{J-1} \sum_{n=0}^{M-1} s_{j,n} \, \delta_{m-Jn-j}$, where $\delta_l$ is the Kronecker delta function. Taking $z$-transforms on both sides of the equality leads to

$$\widetilde{S}(z) = \sum_{j=0}^{J-1} z^{-j} \sum_{n=0}^{M-1} s_{j,n} z^{-Jn} = \sum_{j=0}^{J-1} z^{-j} S_j(z^J). \tag{48}$$

By substituting (15) into (48) and using the definition (17), we can simplify (48) as

$$\widetilde{S}(z) = C(z^{KJ})\widetilde{G}(z). \tag{49}$$

Finally, since $C(z^{KJ})$ is the $z$-transform of the sequence $\sum_n c_n \, \delta_{m-KJn}$, it follows from (49) that $\widetilde{s}_m = (\sum_n c_n \, \delta_{m-KJn}) * \widetilde{g}_m$, and thus (18).

### B. The CRLB of Binary Sensors

We first compute the Fisher information, defined as $I(c) = \mathbb{E}[-\frac{\partial^2}{\partial c^2} \log \mathcal{L}_{\boldsymbol{b}}(c)]$. Using (22), we get

$$\begin{aligned}
I(c) &= \mathbb{E}\left[ -\frac{\partial^2}{\partial c^2} \Big( K_1 \log p_1(c/K) + (K - K_1) \log p_0(c/K) \Big) \right] \\
&= \mathbb{E}\left[ \frac{K_1 \left( p_0''(c/K) p_1(c/K) + p_0'(c/K)^2 \right)}{K^2 p_1(c/K)^2} - \frac{(K - K_1) \left( p_0''(c/K) p_0(c/K) - p_0'(c/K)^2 \right)}{K^2 p_0(c/K)^2} \right],
\end{aligned} \tag{50}$$

where $p_0'(x) = \frac{d}{dx} p_0(x)$ and $p_0''(x) = \frac{d^2}{dx^2} p_0(x)$ are the first and second order derivative of $p_0(x)$, respectively. In reaching (50), we have also used the fact that $p_1(x) = 1 - p_0(x)$ and thus $p_0'(x) = -\frac{d}{dx} p_1(x)$ and $p_0''(x) = -\frac{d^2}{dx^2} p_1(x)$.

Note that $K_1 = \sum_{0 \le m < K} b_m$ is a binomial random variable, and thus its mean can be computed as

$$\mathbb{E}[K_1] = K p_1(c/K) = K(1 - p_0(c/K)).$$

On substituting the above expression into (50), the Fisher information can be simplified as

$$\begin{aligned}
I(c) &= \frac{p_0''(c/K) p_1(c/K) + p_0'(c/K)^2}{K p_1(c/K)} - \frac{p_0''(c/K) p_0(c/K) - p_0'(c/K)^2}{K p_0(c/K)} \\
&= \frac{p_0'(c/K)^2}{K p_0(c/K) p_1(c/K)}.
\end{aligned} \tag{51}$$

Using the definition of $p_0(x)$ in (12), the derivative in the numerator of (51) can be computed as

$$p_0'(x) = -e^{-x} \frac{x^{q-1}}{(q-1)!}. \tag{52}$$

Finally, since $\text{CRLB}_{\text{bin}}(K, q) = 1/I(c)$, we reach (23) by substituting (12) and (52) into (51), and after some straightforward manipulations.

### C. The CRLB of Ideal Unquantized Sensors

Without quantization, the sensor measurements are $\boldsymbol{y} \overset{\text{def}}{=} [y_0, y_1, \ldots, y_{K-1}]^T$, *i.e.*, the number of photons collected at each pixel. The likelihood function in this case is

$$
\begin{aligned}
\mathcal{L}_{\boldsymbol{y}}(c) &\overset{\text{def}}{=} \mathbb{P}(Y_m = y_m, 0 \le m < K; c), \\
&= \prod_{m=0}^{K-1} \frac{(c/K)^{y_m} e^{-c/K}}{y_m!},
\end{aligned}
\tag{53}
$$

where (53) follows from the independence of $\{Y_m\}$ and the expressions (11) and (19).

Computing the Fisher information $I(c) = \mathbb{E}[-\frac{\partial^2}{\partial c^2} \log \mathcal{L}_{\boldsymbol{y}}(c)]$ in this case, we get

$$
\begin{aligned}
I(c) &= \mathbb{E}\left[-\frac{\partial^2}{\partial c^2} \sum_{m=0}^{K-1} \big(y_m \log(c/K) - c/K - \log(y_m!)\big)\right] \\
&= \mathbb{E}\Big[\sum_{m=0}^{K-1} y_m\Big]/c^2.
\end{aligned}
\tag{54}
$$

Since $\{y_m\}$ are drawn from Poisson distributions as in (11), we have $\mathbb{E}[y_m] = s_m = c/K$ for all $m$. It then follows from (54) that $I(c) = K(c/K)/c^2 = 1/c$, and therefore $\text{CRLB}_{\text{ideal}}(K) = 1/I(c) = c$.

### D. Proof of Proposition 3

Using (24), (51) and (52), and through a change of variables $c/K \to x$, we have

$$
\text{CRLB}_{\text{bin}}(K, q)/\text{CRLB}_{\text{ideal}}(K) = \frac{p_0(x) p_1(x)}{x^{2q-1} e^{-2x}/(q-1)!^2}.
\tag{55}
$$

It follows from the properties of incomplete gamma functions that $p_0(x) = \frac{1}{(q-1)!} \int_x^\infty t^{q-1} e^{-t} \, dt$ and $p_1(x) = \frac{1}{(q-1)!} \int_0^x t^{q-1} e^{-t} \, dt$. Using a change of variables $t \to \frac{x^2}{t}$, we can further rewrite $p_0(x)$ as $p_0(x) = \frac{1}{(q-1)!} \int_0^x \left(\frac{x^2}{t}\right)^q e^{-\frac{x^2}{t}} \frac{dt}{t}$. It follows that

$$
\begin{aligned}
\frac{p_0(x) p_1(x)}{x^{2q-1} e^{-2x}/(q-1)!^2} &= \frac{\left(\frac{1}{(q-1)!} \int_0^x \left(\frac{x^2}{t}\right)^q e^{-\frac{x^2}{t}} \frac{dt}{t}\right) \left(\frac{1}{(q-1)!} \int_0^x t^{q-1} e^{-t} \, dt\right)}{x^{2q-1} e^{-2x}/(q-1)!^2} \\
&= x e^{2x} \int_0^x t^{-q-1} e^{-\frac{x^2}{t}} \, dt \int_0^x t^{q-1} e^{-t} \, dt \\
&\ge x e^{2x} \left(\int_0^x t^{-1} e^{-\frac{x^2}{2t} - \frac{t}{2}} \, dt\right)^2,
\end{aligned}
\tag{56}
$$

where in reaching (56) we have used the Cauchy-Schwarz inequality.

It is easy to verify (through a change of variables $t \to \frac{x^2}{t}$) that $\int_0^x t^{-1} e^{-\frac{x^2}{2t} - \frac{t}{2}} \, dt = \int_x^\infty t^{-1} e^{-\frac{x^2}{2t} - \frac{t}{2}} \, dt$. Consequently, the term on the right-hand side of (56) is equal to $x e^{2x} \left(\frac{1}{2} \int_0^\infty t^{-1} e^{-\frac{x^2}{2t} - \frac{t}{2}} \, dt\right)^2$. Through a change of variables $t \to t^2$, we have,

$$
\begin{aligned}
xe^{2x} \left( \frac{1}{2} \int_0^\infty t^{-1} e^{-\frac{x^2}{2t} - \frac{t}{2}} \, dt \right)^2 &= xe^{2x} \left( \int_0^\infty t^{-1} e^{-\frac{x^2}{2t^2} - \frac{t^2}{2}} \, dt \right)^2 \\
&= \left( \sqrt{x} \int_0^\infty t^{-1} e^{-\frac{1}{2} \left( \frac{x}{t} - t \right)^2} \, dt \right)^2 \\
&= \left( \int_{-\infty}^\infty \frac{e^{-\frac{1}{2} u^2}}{\sqrt{4 + u^2/x}} \, du \right)^2,
\end{aligned}
\tag{57}
$$

where (57) is obtained through another change of variables $t \to \frac{u}{2} + \sqrt{x + \frac{u^2}{4}}$.

We can easily verify that (57) is a monotonically increasing function with respect to $x > 0$. So, for $x \geq 1$, (57) is greater than

$$
\left( \int_{-\infty}^\infty e^{-\frac{1}{2} u^2} / \sqrt{4 + u^2} \, du \right)^2 \approx 1.31.
\tag{58}
$$

For $0 \leq x < 1$, we can obtain the following inequalities by keeping the first two terms in $p_0(x)$ and the first term in $p_1(x)$:

$$
\frac{p_0(x) p_1(x)}{x^{2q-1} e^{-2x} / (q-1)!^2} \geq \frac{(1+x) e^{-2x} x^q / q!}{x^{2q-1} e^{-2x} / (q-1)!^2} = \frac{(q-1)!(x^{1-q} + x^{2-q})}{q} \geq \frac{2(q-1)!}{q}.
$$

It is easy to see that $\frac{2(q-1)!}{q}$ is a monotonically increasing function with respect to $q \geq 2$. Therefore,

$$
\frac{2(q-1)!}{q} \geq \frac{4}{3} \approx 1.33 \quad \text{for } q \geq 3.
\tag{59}
$$

Finally, for $q = 2$, we keep the first two terms in $p_0(x)$ and $p_1(x)$ and get

$$
\frac{p_0(x) p_1(x)}{x^{2q-1} e^{-2x} / (q-1)!^2} \geq \frac{(1+x) e^{-2x} \left( x^2/2 + x^3/6 \right)}{x^3 e^{-2x}} = x/6 + 2/3 + \frac{1}{2x} \geq 1.33 \quad \text{for } 0 \leq x \leq 1.
\tag{60}
$$

Combining (58), (59) and (60), we reach the desired result in (26).

Finally, we show that, for $q \geq 2$, the "gap" between $\text{CRLB}_{\text{bin}}(K, q)$ and $\text{CRLB}_{\text{ideal}}(K)$ will only get bigger as the oversampling factor $K$ grows. To that end, we notice that when $K \to \infty$, the variable $x = c/K \to 0$. Keeping the first terms in $p_0(x)$ and $p_1(x)$, we have

$$
\frac{p_0(x) p_1(x)}{x^{2q-1} e^{-2x} / (q-1)!^2} \geq \frac{e^{-2x} x^q / q!}{x^{2q-1} e^{-2x} / (q-1)!^2} = \frac{(q-1)! x^{1-q}}{q}.
$$

For $q \geq 2$, the above quantity goes to infinity as $x \to 0$. Therefore,

$$
\lim_{K \to \infty} \text{CRLB}_{\text{bin}}(K, q) / \text{CRLB}_{\text{ideal}}(K) = \infty.
$$

### E. Proof of Theorem 1

When $q = 1$, we have $p_0(x) = e^{-x}$ and thus $p_0^{[-1]}(x) = -\log(x)$. In this case, the MLE solution in (28) can be rewritten as

$$\widehat{c}_{\text{ML}}(\boldsymbol{b}) = \begin{cases} -K \log(1 - K_1/K), & \text{if } 0 \leq K_1 \leq K(1 - e^{-S/K}), \\ S, & \text{otherwise.} \end{cases}$$

We note that $-K \log(1 - K_1/K) = K_1 + \frac{K_1^2}{2K} + \frac{K_1^3}{3K^2} + \dots$ and that $\lim_{K \to \infty} K(1 - e^{-S/K}) = S$. Thus, for sufficiently large $K$, the above MLE solution can be further simplified as

$$\widehat{c}_{\text{ML}}(\boldsymbol{b}) = \begin{cases} K_1 + \mathcal{O}(\frac{1}{K}), & \text{if } 0 \leq K_1 < S, \\ S, & \text{otherwise.} \end{cases} \tag{61}$$

Without loss of generality, we assume that $S$ is an integer in what follows. The expected value of the MLE then becomes

$$\mathbb{E}[\widehat{c}_{\text{ML}}(\boldsymbol{b})] = \sum_{n=0}^{S-1} n \, \mathbb{P}(K_1 = n) + S \sum_{n=S}^{K} \mathbb{P}(K_1 = n) + \mathcal{O}(1/K).$$

Using the following identity $c = \sum_{n=0}^{\infty} n \frac{c^n e^{-c}}{n!}$ about the mean of a Poisson random variable, we have

$$\left| \mathbb{E}[\widehat{c}_{\text{ML}}(\boldsymbol{b})] - c \right| = \left| \sum_{n=0}^{S-1} n \left( \mathbb{P}(K_1 = n) - \frac{c^n e^{-c}}{n!} \right) + S \sum_{n=S}^{K} \mathbb{P}(K_1 = n) - \sum_{n=S}^{\infty} n \frac{c^n e^{-c}}{n!} + \mathcal{O}(1/K) \right|$$

$$\leq S \left| \sum_{n=0}^{S-1} \left( \mathbb{P}(K_1 = n) - \frac{c^n e^{-c}}{n!} \right) \right| + S \sum_{n=S}^{K} \mathbb{P}(K_1 = n) + \sum_{n=S}^{\infty} \frac{c^n e^{-c}}{(n-1)!} + \mathcal{O}(1/K). \tag{62}$$

In what follows, we derive bounds for the quantities on the right-hand side of the above inequality. First, consider the probability $\mathbb{P}(K_1 = n)$. Since $K_1$ is a binomial random variable, we have

$$\mathbb{P}(K_1 = n) = \binom{K}{n} (1 - p_0(c/K))^n p_0(c/K)^{(K-n)}$$

$$= \frac{K(K-1) \dots (K-n+1)}{n! K^n} (K(1 - e^{-c/K}))^n e^{-c(K-n)/K}. \tag{63}$$

For every $n < S$, it is easy to verify that $\frac{K(K-1)\dots(K-n+1)}{K^n} = 1 + \mathcal{O}(1/K)$, $(K(1 - e^{-c/K}))^n = c^n + \mathcal{O}(1/K)$ and $e^{-c(K-n)/K} = e^{-c} + \mathcal{O}(1/K)$. Thus, for any $n < S$, we can simplify (63) as

$$\mathbb{P}(K_1 = n) = \frac{c^n}{n!} e^{-c} + \mathcal{O}(1/K). \tag{64}$$

It follows that

$$S \left| \sum_{n=0}^{S-1} \left( \mathbb{P}(K_1 = n) - \frac{c^n e^{-c}}{n!} \right) \right| = \mathcal{O}(1/K). \tag{65}$$

Next, consider the second term on the right-hand side of (62).

$$S \sum_{n=S}^{K} \mathbb{P}(K_1 = n) = S(1 - \sum_{n=0}^{S-1} \mathbb{P}(K_1 = n))$$

$$= S(1 - \sum_{n=0}^{S-1} \frac{c^n}{n!} e^{-c}) + \mathcal{O}(1/K) \tag{66}$$

$$= S \sum_{n=S}^{\infty} \frac{c^n}{n!} e^{-c} + \mathcal{O}(1/K)$$

$$\leq S e^{-c} \left( \frac{ec}{S} \right)^S + \mathcal{O}(1/K), \tag{67}$$

for all $c < S$, where (66) follows from (64) and the inequality (67) is due to the Chernoff bound on the tail of Poisson distributions [24]. Similarly, the third term on the right-hand side of (62) can be rewritten as

$$\sum_{n=S}^{\infty} \frac{c^n e^{-c}}{(n-1)!} = c \sum_{n=S-1}^{\infty} \frac{c^n e^{-c}}{n!} \leq c \, e^{-c} \left( \frac{ec}{S-1} \right)^{S-1}, \tag{68}$$

where the inequality is again an application of the Chernoff bound. Finally, on substituting (65), (67) and (68) into (62), and after some simple manipulations, we reach (29).

The proof for the mean-squared error formula (30) is similar. Using (61), we have

$$\mathbb{E}\left[ (\widehat{c}_{\mathrm{ML}}(\boldsymbol{b}) - c)^2 \right] = \sum_{n=0}^{S-1} (n-c)^2 \mathbb{P}(K_1 = n) + (S-c)^2 \sum_{n=S}^{K} \mathbb{P}(K_1 = n) + \mathcal{O}(1/K)$$

$$= \sum_{n=0}^{S-1} (n-c)^2 \frac{c^n e^{-c}}{n!} + (S-c)^2 \sum_{n=S}^{\infty} \frac{c^n e^{-c}}{n!} + \mathcal{O}(1/K), \tag{69}$$

where in reaching (69), we have used the estimation (64) of the Binomial probabilities. We note that the variance of a Poisson random variable is equal to its mean. Thus, $c = \sum_{n=0}^{\infty} (n-c)^2 \frac{c^n e^{-c}}{n!}$. On combining this identity with (69),

$$\left| \mathbb{E}\left[ (\widehat{c}_{\mathrm{ML}}(\boldsymbol{b}) - c)^2 \right] - c \right| = \left| (S-c)^2 \sum_{n=S}^{\infty} \frac{c^n e^{-c}}{n!} - \sum_{n=S}^{\infty} (n-c)^2 \frac{c^n e^{-c}}{n!} + \mathcal{O}(1/K) \right|$$

$$\leq 2 \sum_{n=S}^{\infty} n^2 \frac{c^n e^{-c}}{n!} + \mathcal{O}(1/K), \qquad \text{for } c \leq S$$

$$\leq 2c(c+1) \sum_{n=S-2}^{\infty} \frac{c^n e^{-c}}{n!} + \mathcal{O}(1/K), \qquad \text{for } c < S - 2.$$

Applying the Chernoff bound to the above inequality, we get (30).

### F. Proof of Proposition 4

We have $p_0(x) = e^{-x} \sum_{n=0}^{q-1} \frac{x^n}{n!}$, and thus $1 - p_0(x) = e^{-x} \sum_{n=q}^{\infty} \frac{x^n}{n!}$. It follows that

$$
\begin{aligned}
K(1 - p_0(S/K)) &= K e^{-S/K} \left( \frac{S^q}{K^q q!} + \frac{S^{q+1}}{K^{q+1}(q+1)!} + \dots \right) \\
&= e^{-S/K} \left( \frac{S^q}{K^{q-1} q!} + \frac{S^{q+1}}{K^q (q+1)!} + \dots \right).
\end{aligned}
$$

For $q \geq 2$ and any fixed constant $S$, the above quantity converges to 0 as $K$ tends to infinity. As a result, for sufficiently large $K$, the MLE solution in (28) can be simplified as

$$
\widehat{c}_{\text{ML}}(\boldsymbol{b}) = \begin{cases} 0, & \text{if } K_1 = 0, \\ S, & \text{otherwise,} \end{cases} \tag{70}
$$

where we have also used the fact that $p_0(0) = 1$ and thus $p_0^{[-1]}(1) = 0$. Using (70), we can compute the expected value of the MLE as

$$
\mathbb{E}[\widehat{c}_{\text{ML}}(\boldsymbol{b})] = 0 \, \mathbb{P}(K_1 = 0) + S(1 - \mathbb{P}(K_1 = 0)) = S(1 - \mathbb{P}(K_1 = 0)). \tag{71}
$$

We have $K_1 = 0$ when all the pixel responses are uniformly 0. The probability of seeing such an event is

$$
\mathbb{P}(K_1 = 0) = p_0(c/K)^K = e^{-c} \left( 1 + \frac{c}{K} + \dots + \frac{c^{q-1}}{K^{q-1}(q-1)!} \right)^K,
$$

which converges to 1 as $K$ tends to infinity, *i.e.*,

$$
\lim_{K \to \infty} \mathbb{P}(K_1 = 0) = 1. \tag{72}
$$

Substituting (72) into (71), we get (31).

Next, we compute the MSE as

$$
\mathbb{E}\left[ (\widehat{c}_{\text{ML}}(\boldsymbol{b}) - c)^2 \right] = c^2 \, \mathbb{P}(K_1 = 0) + (S - c)^2 \left( 1 - \mathbb{P}(K_1 = 0) \right),
$$

which, upon taking the limit as $K \to \infty$, leads to (32).

### G. Proof of Lemma 2

The function $h(x) \stackrel{\text{def}}{=} \log \sum_{k=i}^{j} \frac{x^k e^{-x}}{k!}$ is continuously differentiable on the interval $(0, \infty)$. Therefore, to establish its concavity, we just need to show that its second derivative is nonpositive. To that end, we first introduce a sequence of functions $\{r_k(x)\}_{k \in Z \cup \{\infty\}}$, defined as

$$
r_k(x) \stackrel{\text{def}}{=} \begin{cases} x^k/k!, & \text{if } 0 \leq k < \infty; \\ 0, & \text{if } k < 0 \text{ or } k = \infty. \end{cases} \tag{73}
$$

It is straightforward to verify that $\frac{d}{dx}r_k(x) = r_{k-1}(x)$ for all $k \in Z \cup \{\infty\}$. Now, rewriting $h(x)$ as $\log \sum_{k=i}^{j} r_k(x) - x$ and computing its second derivative, we get

$$\frac{d^2}{dx^2}h(x) = \frac{(\sum_{k=i}^{j} r_{k-2})(\sum_{k=i}^{j} r_k) - (\sum_{k=i}^{j} r_{k-1})^2}{(\sum_{k=i}^{j} r_k)^2}, \tag{74}$$

where we have omitted the function argument $x$ in $r_k(x), r_{k-1}(x)$ and $r_{k-2}(x)$ for notational simplicity.

Recall that our goal is to show that $\frac{d^2}{dx^2}h(x) \leq 0$, for $x > 0$. Since the denominator of (74) is always positive, we just need to focus on its numerator. Using the identities $\sum_{i \leq k \leq j} r_k = \sum_{i \leq k \leq j} r_{k-1} + r_j - r_i$ and $\sum_{i \leq k \leq j} r_{k-1} = \sum_{i \leq k \leq j} r_{k-2} + r_{j-1} - r_{i-1}$, we can simplify the numerator of (74) as follows:

$$\Big( \sum_{i \leq k \leq j} r_{k-2} \Big)\Big( \sum_{i \leq k \leq j} r_{k-1} + r_j - r_i \Big) - \Big( \sum_{i \leq k \leq j} r_{k-1} \Big)\Big( \sum_{i \leq k \leq j} r_{k-2} + r_{j-1} - r_{i-1} \Big)$$

$$= \sum_{i \leq k \leq j} \Big( (r_{k-2}r_j - r_{k-1}r_{j-1}) + (r_{k-1}r_{i-1} - r_{k-2}r_i) \Big). \tag{75}$$

In what follows, we show that

$$r_{k-2}(x)r_j(x) - r_{k-1}(x)r_{j-1}(x) \leq 0 \tag{76}$$

for arbitrary choices of $x \geq 0$ and $i \leq k \leq j$, where $0 \leq i \leq j < \infty$ or $0 \leq i < j = \infty$. Note that, when $k < 2$ or $j = \infty$, the left-hand side of (76) becomes $-r_{k-1}(x)r_{j-1}(x)$ and thus (76) automatically holds. Now, assume that $k \geq 2$ and $j < \infty$. From the definition in (73), the left-hand side of (76) is

$$\frac{x^{k-2}x^j}{(k-2)!j!} - \frac{x^{k-1}x^{j-1}}{(k-1)!(j-1)!} = \frac{x^{k+j-2}}{(k-2)!(j-1)!}\left( \frac{1}{j} - \frac{1}{k-1} \right) \leq 0$$

for $i \leq k \leq j$. Using similar arguments, we can also show that

$$r_{k-1}(x)r_{i-1}(x) - r_{k-2}(x)r_i(x) \leq 0, \qquad \text{for } x \geq 0. \tag{77}$$

On substituting the inequalities (76) and (77) into (75), we verify that the numerator of (74) is nonpositive, and therefore $\frac{d^2}{dx^2}h(x) \leq 0$, for all $x > 0$.

### H. Proof of Proposition 5

Expressing the signal processing operations in Fig. 6 in the $z$-domain, we have

$$U(z) = A(z^K)G(z)$$

$$= A(z^K)G_0(z^K) + z^{-1}A(z^K)G_1(z^K) + \ldots + z^{-(K-1)}A(z^K)G_{K-1}(z^K), \tag{78}$$

where $A(z^K)$ in the first equality is the $z$-transform of the $K$-times upsampled version of $a_m$, and (78) follows from (45). Similar to (45), we can expand $U(z)$ in terms of the $z$-transforms of its polyphase components, as

$$U(z) = U_0(z^K) + z^{-1}U_1(z^K) + \ldots + z^{-(K-1)}U_{K-1}(z^K). \tag{79}$$

Comparing (78) against (79) and using the uniqueness of the polyphase decomposition, we conclude that $U_k(z) = A(z)G_k(z)$, for all $0 \le k < K$.

Now, consider Fig. 6(b). We note that the $z$-transform of $g_{-m}$ is $G(z^{-1})$. Denote by $d_m$ the output of the filtering operation. Then, its $z$-transform can be computed as

$$
\begin{aligned}
D(z) &= B(z)G(z^{-1}) \\
&= \left( \sum_{k=0}^{K-1} z^{-k} B_k(z^K) \right) \left( \sum_{k=0}^{K-1} z^k G_k(z^{-K}) \right) \\
&= \sum_{k=0}^{K-1} B_k(z^K)G_k(z^{-K}) + \sum_{0 \le i \ne j < K} z^{j-i} B_i(z^K)G_j(z^K).
\end{aligned}
\tag{80}
$$

When downsampling $d_m$ by $K$, only the first term on the right-hand side of (80) will be retained; the second term is "invisible" to the sampling operation due to mismatched supports. Therefore, we have, after downsampling, $V(z) = \sum_{k=0}^{K-1} B_k(z)G_k(z^{-1})$.

## REFERENCES

[1] T. H. James, *The Theory of the Photographic Process*, 4th ed. New York: Macmillan Publishing Co., Inc., 1977.

[2] S. A. Ciarcia, "A 64K-bit dynamic RAM chip is the visual sensor in this digital image camera," *Byte Magazine*, pp. 21–31, Sep. 1983.

[3] Y. K. Park, S. H. Lee, J. W. Lee *et al.*, "Fully integrated 56 nm DRAM technology for 1 Gb DRAM," in *IEEE Symposium on VLSI Technology*, Kyoto, Japan, Jun. 2007.

[4] J. C. Candy and G. C. Temes, *Oversampling Delta-Sigma Data Converters — Theory, Design and Simulation*. New York, NY: IEEE Press, 1992.

[5] V. K. Goyal, M. Vetterli, and N. T. Thao, "Quantized overcomplete expansions in $\mathbb{R}^n$: Analysis, synthesis and algorithms," *IEEE Trans. Inf. Theory*, vol. 44, no. 1, pp. 16–31, Jan. 1998.

[6] P. T. Boufounos and A. V. Oppenheim, "Quantization noise shaping on arbitrary frame expansions," *EURASIP J. Applied Signal Processing*, pp. 1–12, Jan. 2006.

[7] Z. Cvetković and I. Daubechies, "Single-Bit oversampled A/D conversion with exponential accuracy in the bit rate," *IEEE Trans. Inf. Theory*, vol. 53, no. 11, pp. 3979–3989, Nov. 2007.

[8] E. R. Fossum, "What to do with sub-diffraction-limit (SDL) pixels? – A proposal for a gigapixel digital film sensor (DFS)," in *IEEE Workshop on Charge-Coupled Devices and Advanced Image Sensors*, Nagano, Jun. 2005, pp. 214–217.

[9] M. Unser and M. Eden, "Maximum likelihood estimation of linear signal parameters for Poisson processes," *IEEE Trans. Acoust., Speech, and Signal Process.*, vol. 36, no. 6, pp. 942–945, Jun. 1988.

[10] K. E. Timmermann and R. D. Nowak, "Multiscale modeling and estimation of Poisson processes with application to photon-limited imaging," *IEEE Trans. Inf. Theory*, vol. 45, no. 3, pp. 846–862, Apr. 1999.

[11] R. D. Nowak and E. D. Kolaczyk, "A statistical multiscale framework for Poisson inverse problems," *IEEE Trans. Inf. Theory*, vol. 46, no. 5, pp. 1811–1825, Aug. 2000.

[12] R. M. Willet and R. D. Nowak, "Platelets: a multiscale approach for recovering edges and surfaces in photon-limited medical imaging," *IEEE Trans. Med. Imag.*, vol. 22, no. 3, pp. 332–350, Mar. 2003.

[13] R. M. Willett and R. D. Nowak, "Multiscale Poisson intensity and density estimation," *IEEE Trans. Inf. Theory*, vol. 53, no. 9, pp. 3171–3187, Sep. 2007.

[14] H. V. Poor, *An Introduction to Signal Detection and Estimation*, 2nd ed.   New York, NY: Springer-Verlag, 1994.

[15] P. P. Vaidyanathan, *Multirate Systems and Filter Banks*.   Englewood Cliffs, NJ: Prentice-Hall, 1993.

[16] M. Vetterli and J. Kovačević, *Wavelets and Subband Coding*.   Englewood Cliffs, NJ: Prentice Hall, 1995. [Online]. Available: http://www.waveletsandsubbandcoding.org/

[17] L. Carrara, C. Niclass, N. Scheidegger, H. Shea, and E. Charbon, "A gamma, X-ray and high energy proton radiation-tolerant CMOS image sensor for space applications," in *IEEE International Solid-State Circuits Conference*, Feb. 2009, pp. 40–41.

[18] M. Born and E. Wolf, *Principles of Optics*, 7th ed.   Cambridge: Cambridge University Press, 1999.

[19] K. Fife, A. E. Gamal, and H.-S. P. Wong, "A multi-aperture image sensor with $0.7\mu$m pixels in $0.11\mu$m CMOS technology," *IEEE Journal of Solid-State Circuits*, vol. 43, no. 12, pp. 2990–3005, Dec. 2008.

[20] M. Unser, "Splines: A perfect fit for signal and image processing," *IEEE Signal Process. Mag.*, pp. 22–38, Nov. 1999.

[21] H. Wakabayashi, K. Yamaguchi, M. Okano *et al.*, "A 1/2.3-inch 10.3Mpixel 50frame/s black-illuminated CMOS image sensor," in *Proc. IEEE International Solid-State Circuits Conference*, Feb. 2010, pp. 410–411.

[22] P. E. Debevec and J. Malik, "Recovering high dynamic range radiance maps from photographs," in *Proc. 24th Annual Conference on Computer Graphics and Interactive Techniques*, Los Angeles, Aug. 1997, pp. 369–378.

[23] E. Reinhard, G. Ward, S. Pattanaik, and P. Debevec, *High dynamic range imaging: Acquisition, display and image-based lighting*.   San Francisco, CA: Morgan Kaufmann, 2006.

[24] T. Hagerup and C. Rüb, "A guided tour of Chernoff bounds," *Information Processing Letters*, vol. 33, no. 6, pp. 305–308, Feb. 1990.